\documentclass{elsart}
\usepackage{mathptmx}
\usepackage{bm,amsfonts}
\makeatletter
\@ifundefined{pdfoutput} 
{ 
\usepackage[dvips]{graphicx,color}
}
{ 
\ifnum\pdfoutput=0\relax
\usepackage[dvips]{graphicx,color}
\fi
\ifnum\pdfoutput=1\relax
\usepackage[pdftex]{graphicx,color}
\fi
}
\makeatother
\newcommand{\beq}{\begin{equation}}
\newcommand{\eeq}{\end{equation}}
\newcommand{\beqa}{\begin{eqnarray}}
\newcommand{\eeqa}{\end{eqnarray}}
\newcommand{\nn}{\nonumber \\}
\newcommand {\np}[1]{{\mbox{\textrm{:}\,}{#1}{\,\textrm{:}}} }
\def \e {\mathrm{e}}

\def \la {\langle}
\def \ra {\rangle}
\def \s {\sigma}

\def \B {{\mathcal B}}
\def \C {{\mathbb C}}

\def \I {{\mathbb I}}

\def \Z {{\mathbb Z}}

\def \qh {\mathrm{qh}}

\def \tx {\tilde{x}}

\def \D {\Delta}
\def \Pf {\mathrm{Pf}}

\def \Re {\mathrm{Re} \, }%

\def \H {{\mathcal H}}

\begin{document}
\begin{frontmatter}
\title{Towards a universal set of topologically protected gates for
	quantum computation with Pfaffian qubits}
\author{Lachezar S. Georgiev}
\ead{lgeorg@inrne.bas.bg}
\address{Institute for Nuclear Research and
Nuclear Energy \\
 Tsarigradsko Chaussee 72,  1784 Sofia, BULGARIA}
\begin{abstract}
We review the topological quantum computation scheme of Das Sarma et al.
from the perspective of the conformal field theory for the
two-dimensional critical Ising model. This scheme originally used the
\textit{monodromy}
properties of the non-Abelian excitations in the Pfaffian quantum Hall state
to construct elementary qubits and execute logical NOT on them.
We extend the scheme of Das Sarma et al. by exploiting the explicit
\textit{braiding}
transformations for the Pfaffian wave functions containing 4 and 6 quasiholes
 to implement, for the
first time in this context,  the single-qubit Hadamard and phase gates and
the two-qubit
Controlled-NOT gate over Pfaffian qubits in a topologically protected way.
In more detail, we  explicitly construct the unitary representations of the
braid groups $\B_4$, $\B_6$ and $\B_8$ and use the elementary braid matrices 
to implement one-, two- and three-qubit gates.
We also propose to construct a topologically protected Toffoli gate,
in terms of a braid-group based Controlled-Controlled-Z gate
 precursor.
Finally we discuss some difficulties arising in the embedding of the Clifford
gates and address several important questions about topological quantum 
computation in general.
\end{abstract}
\end{frontmatter}
\section{Introduction}
In contrast to the three dimensional world, where we could only find
bosons and  fermions, the statistics of localized objects in two
dimensions turned out to be much richer, and includes fractional or
anyonic  statistics \cite{wilczek}. In the simplest case of Abelian
fractional statistics the counter-clockwise exchange of two anyons
multiplies the many-body state by the statistical phase $\exp(i\theta)$,
where $\theta/\pi$ might be a fractional number. The most important
distinction between anyons on one side and bosons or fermions on the other
 is that the clockwise and  counter-clockwise exchanges are significantly
different so that the many-body wave functions belong to the representations
spaces not of the permutation group but of the braid group
\cite{wilczek,birman}.
When the dimension of the braid-group representation
is bigger  than 1  the corresponding (quasi)particles are called plektons
\cite{fredenhagen-rehren-shroer,froehlich-gabbiani,fredenhagen-gaberdiel},
or non-Abelian anyons, and the exchange of two such anyons
results in a non-trivial statistical matrix acting over degenerate space of
many-body states. As the matrices representing different exchanges do not
commute in general, this kind of quasiparticle statistics is called
non-Abelian.

The most promising two-dimensional system, in which  non-Abelian statistics
may  eventually  be observed seems to be the fractional quantum Hall (FQH)
state in
the second Landau level with filling factor $\nu=5/2$, which is now routinely
observed in ultra high-mobility samples \cite{eisen2002,xia}.
Convincing analytical \cite{read-green,ivanov} and numerical \cite{morf,5-2hr}
evidence suggest that
this state is most likely in the universality class of the Pfaffian FQH
state constructed by Moore and Read \cite{mr} using correlation functions of
certain operators in an appropriate $1+1$ dimensional conformal field
theory (CFT) \cite{CFT-book} including the Ising model \cite{mr,CMP99}.
Experimental tests of fractional statistics appear to be
much more difficult than those  which confirmed the
fractional electric charge \cite{picciotto} of the FQH quasiparticles.
However, it turns out that consequences of the presence of
non-Abelian excitations might in fact be easier to observe than the Abelian
fractional statistics itself despite the structural complications.
In recent theoretical work interesting proposals for detection of the
 non-Abelian statistics of the quasiparticles in the $\nu=5/2$ FQH state
\cite{stern-halperin,bonderson-5-2,5-2AB,feldman-kitaev} and in the
$\nu=12/5$ FQH state \cite{bonderson-12-5,chung-stone}, which is expected to
be in the universality class of the $k=3$ parafermion Hall state
\cite{rr,NPB2001},   have been made.

It is quite remarkable that in addition to its fundamental significance,
the non-Abelian quantum statistics might become practically important
for quantum computation \cite{nielsen-chuang}.
Although the ideas behind quantum information processing
are simply based on the well-established fundamental postulates of the
quantum theory, its
exponentially growing computational power could not have been
exploited  so far due to the unavoidable effects of quantum noise and
decoherence as  a result of interaction of the qubits with their environment.
Even the quantum error-correcting algorithms
\cite{nielsen-chuang}, which allow to use operations containing
certain level of errors, could not help creating a quantum computer with
more than a few qubits.
In this context recently appeared the idea of
\textit{topological quantum computation}  (TQC)
\cite{kitaev-TQC,kauffman-braid,preskill-TQC}.
Because the interactions
leading to decoherence are presumably local we can try to
avoid them
by encoding quantum information non-locally, using some global
e.g., topological characteristics of the system. This
\textit{topological protection of qubit operations} means that
quantum information is inaccessible to local interactions, because they
cannot distinguish between the computational basis states and hence cannot
lead to decoherence and noise
\cite{kitaev-TQC,preskill-TQC,sarma-freedman-nayak,freedman-nayak-walker}.
That is why topological gates are believed to be exact operations, which
might potentially allow for the topologically protected quantum computation.

The FQH liquid is a perfect candidate   for TQC
because it possesses a number of topological properties which are universal,
i.e., robust against the variations of the interactions details,
and could be successfully described by rational conformal field (RCFT)
theories \cite{fro-stu-thi,fro-king,fro2000}, which capture the universality
classes of the FQH edge
excitations (see also \cite{5-2} and Refs. therein).
The main idea is to use the braiding matrices
\cite{fredenhagen-rehren-shroer,froehlich-gabbiani,preskill-TQC}
corresponding to the
exchanges of FQH quasiparticles, acting over a degenerate set of
topological many-body states,   to implement arbitrary unitary transformations
\cite{kitaev-TQC,kauffman-braid,preskill-TQC,werner-kauffman}.
Because the single qubit space is two-dimensional,  we need degenerate
spaces of quasiparticle correlation functions with dimension at least 2.
While the Abelian FQH quasiparticles have degenerate spaces on non-trivial
manifolds such as torus \cite{kitaev-TQC,martin-delgado-TQC}, such
constructions are not appropriate for experimental realization in planar
systems, such  as the FQH liquids. Alternatively,  by Abelian anyons, one
could realize in the plane only diagonal gates, such as Controlled-phase
gates \cite{averin-goldman} but cannot implement non-diagonal gates such as
 the Hadamard gate.
In contrast, the non-Abelian FQH quasiparticles by definition
have degenerate spaces even in planar geometry, which explains why
the non-Abelian anyons might be more appropriate  for TQC.
The only residual source of noise and decoherence is due to thermally
activated quasiparticle--quasihole pairs, which might execute
unwanted braidings. Fortunately, these processes  are exponentially
suppressed at low temperature by the bulk energy gap, which leads to
astronomical precision of quantum information processing.

In a recent paper \cite{sarma-freedman-nayak}
Das Sarma et al. proposed to use the expected non-Abelian  statistics of the
quasiparticles in the Pfaffian FQH state
to construct an elementary qubit and execute a logical NOT gate on it.
The NOT-gate construction is very important for quantum computation because it
underlies both the construction of the single-qubit gates and of the
Controlled-NOT (CNOT) gate \cite{nielsen-chuang}.
However the NOT gate is certainly not sufficient for universal computation,
though it is reversible,  because the universal classical gate,
NAND/NOT, built from NOT is irreversible \cite{preskill-QC},
while all quantum gates must be reversible.
Therefore, in the TQC scheme of Das Sarma et al.,  we need to implement
the CNOT gate (or reversible XOR), which plays a central role
in the universal quantum computer \cite{nielsen-chuang,preskill-QC}.

The FQH state at $\nu=5/2$ has one big advantage with respect to
TQC---it is the most stable state, i.e., the one with the highest
bulk energy gap, among all FQH states in which non-Abelian quasiparticle
statistics is expected to be realized.
Recent measurements of the energy gap \cite{eisen2002,xia}
suggest that ratio noise/signal is expected to be of the order of
$10^{-30}$ or even lower, which is an unprecedented number in the quantum
computation field and might potentially allow for the construction of a
truly scalable quantum computation platform.
On the other hand, one serious drawback
is that the quasiparticles braiding matrices cannot be used alone for
universal quantum computation because, as we shall see later, the
representation of the braid group turns out to be finite.
This has to be compared with the state
at $\nu=12/5$, whose braiding matrices are expected to be universal, but whose
energy gap is an order of magnitude lower than the $\nu=5/2$ one, which
increases dramatically the noise/signal ratio.

In this paper we extend the TQC scheme of Das Sarma et al.
\cite{sarma-freedman-nayak}, which was originally based on monodromy 
transformations of the Pfaffian wave functions, in such a way
to construct by braiding the single-qubit Hadamard and phase gates 
as well as the two-qubit  Controlled-Z and
Controlled-NOT gates. These constructions are naturally 
topologically protected. In addition we investigate some possibilities
for topologically protected realization of the three-qubit Toffoli gate.

\textit{Summary of results}:
  we review in Sect.~\ref{sec:sarma} the TQC scheme of
Das Sarma et al.  and introduce  in Sect.~\ref{sec:4-qh} the 4-quasihole wave
functions of Ref.~\cite{sarma-freedman-nayak} which we shall use to derive
the elementary exchange matrices that will represent single-qubit gates.
In Sect.~\ref{sec:read-out} we consider in more detail the
read-out transformation for the qubit of Ref.~\cite{sarma-freedman-nayak}
by using the analytic properties of the 4-quasiholes
states and prove, under certain conditions, the conjecture of 
Ref.~\cite{sarma-freedman-nayak} that the 
state of the qubit does not change after transferring one quasihole from antidot 1 
to antidot 2, which is crucial for the construction, initialization and manipulation 
of the Pfaffian qubits. When these conditions are not satisfied, the TQC scheme of
Ref.~\cite{sarma-freedman-nayak} is going to fail.
 We prove in Sect.~\ref{sec:orthogonality}
the orthogonality of the 4-quasiholes states forming the computational basis,
which is very important for their quantum distinguishability.
Deriving explicitly in Sect.~\ref{sec:R4}  the complete set of exchange matrices for the
4-quasiholes states, which has been partially done in Ref.~\cite{nayak-wilczek} and 
completely reproduced in  Ref.~\cite{slingerland-bais} using the underlying 
quantum group structure for the parafermion quantum Hall states, we construct in 
Sect.~\ref{sec:single-qubit} all
single-qubit gates, except for the $\pi/8$ gate, entirely in terms of
quasihole braiding. 
Then, in Sect.~\ref{sec:2qubits} we propose a natural two-qubit construction
in terms of 6-quasiholes states and obtain explicitly the exchange matrices
for these  states. In Sects.~\ref{sec:CZ} and ~\ref{sec:CNOT} we
implement the Controlled-Z and  Controlled-NOT gates
entirely in terms of 6-quasiholes braidings. To the best of the author's 
knowledge
this is the first explicit construction of these gates in the Pfaffian state,
which is exact and topologically protected.
We also construct in Sects.~\ref{sec:g_3} and \ref{sec:non-demolition}
 the Bravyi--Kitaev two-qubit gate $g_3$
and the non-demolition charge measurement  gate, respectively,
 in terms of 6-quasiholes braidings.
While the above gates are not sufficient for universal quantum computation
they are known to form a Clifford group
\cite{bravyi-kitaev-Clifford-QC,martin-delgado-TQC}, which plays an
extremely important role  in error-correcting algorithms and, in particular
could be efficiently
used in such applications as quantum teleportation and super-dense coding
\cite{nielsen-chuang}.
Moreover, if the Clifford group is supplemented by the so-called magic 
states or
 noisy ancillas that could already be used for universal quantum computation
\cite{bravyi-kitaev-Clifford-QC}.
In addition to the Clifford-group gates, instead of using the $\pi/8$
gate, we propose in Sect.~\ref{sec:Toffoli} to implement in a topologically
protected way the Toffoli gate in terms of the CNOT and the
Controlled-S gate, or by a braid-group based Controlled-Controlled-Z gate 
precursor, which would actually form a universal set of topological
protected gates realized with  Pfaffian qubits.
It appears that there is an additional topological entanglement in the
 three-qubit systems defined by 8 Pfaffian quasiholes, which leads to 
complications in the embedding of the one- and two-qubit gates into systems 
with three or more qubits. This phenomenon seems to be common for all 
topological quantum computation schemes based on the braid matrices of 
the non-Abelian FQH anyons.
\section{The TQC scheme of Das Sarma et al.}
\label{sec:sarma}
The main idea of Ref.~\cite{sarma-freedman-nayak,freedman-nayak-walker}
is to use the wave functions
of the Pfaffian FQH state with 4 quasiholes to form an elementary qubit.
Then quantum gates can be executed by braiding some of the quasiholes
(i.e., by counter-clockwise exchanges of quasiholes) leading to unitary
transformations in the qubit space.
When the positions of the quasiholes are fixed these wave functions
form a 2-dimensional space  which could be used as the single-qubit space.
In general, the wave functions
containing $2n$ quasiholes with fixed positions form a linear space with
dimension $2^{n-1}$ \cite{nayak-wilczek}. The main reason for the exponential
 increase of the space dimension is the non-Abelian statistics of the
quasiholes, i.e,
 when two quasiholes are fused together (taken to the same point in the
coordinate plane)
  the result contains  more than one quasiparticle due to the chiral
Ising model fusion rule \cite{mr}
\beq\label{fusion}
\s \times \s = \I +\psi .
\eeq
The exponential degeneracy of the $2n$ quasiholes spaces can be
alternatively understood by interpreting  the Pfaffian state as
a $p$-wave superconductor of composite fermions \cite{read-green},
where the non-Abelian quasiholes are represented by half-quantum vortices,
and their non-Abelian statistics follows from the existence of
Majorana  zero-modes in the vortex cores obtained as solutions of
the Bogolubov-de Gennes equations
\cite{read-green,ivanov,stern-oppen-mariani}.

\begin{figure}[htb]
\centering
\includegraphics*[bb=25 560 562 820,width=\textwidth]{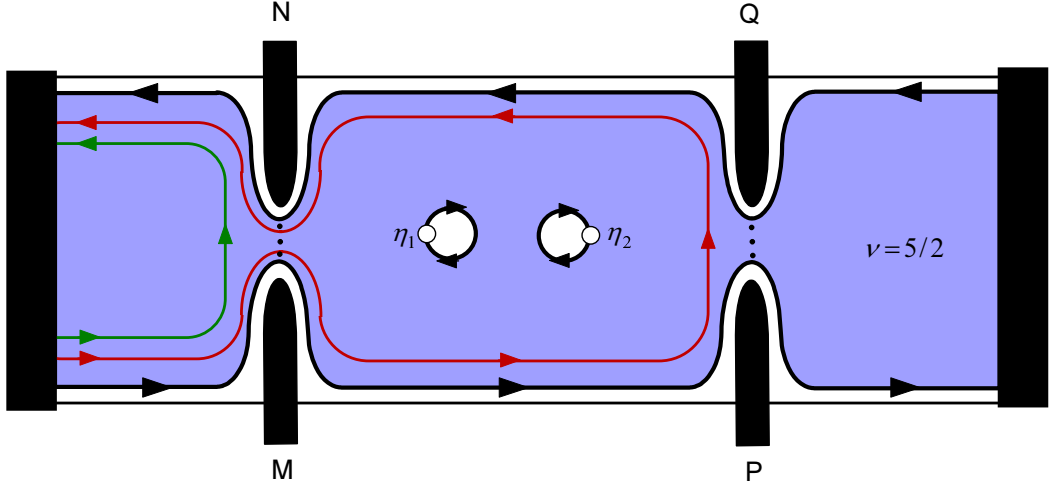}
 \caption{(Color online). A $\nu=5/2$ Hall bar with two antidots,
	on which the
 	quasiholes (denoted symbolically  by small empty circles) with
 	coordinates $\eta_1$ and $\eta_2$ comprising the qubit are localized,
	and 4 front gates, M, N, P and Q creating tunneling  	constrictions.
	Black arrows depict the edge states, while the green and red arrows denote two
	alternative tunneling channels, a direct one and such enclosing the two antidots. }
	\label{fig:qubit}
\end{figure}

One way to keep the positions of the quasiholes fixed is to introduce antidots
\cite{antidot-goldman}
(lithographically defined potential hills expelling the FQH fluid and
creating a ``hole'' or ``island'' inside it) in the FQH liquid and localize
the quasiholes there as shown on Fig.~\ref{fig:qubit}.
The positions of the quasiholes are denoted by $\eta_a$, where $a=1,\ldots,4$,
and we assume that the quasiholes with coordinates $\eta_1$ and  $\eta_2$
 form our qubit, while $\eta_3$ and $\eta_4$ (not shown explicitly on
 Fig.~\ref{fig:qubit}) are used to measure and manipulate the qubit's state
 \cite{sarma-freedman-nayak,nayak-private}.
\subsection{State of the qubit and its initialization}
In order to have a real TQC scheme we need to specialize the computational
basis
$\{|0\ra, \, |1\ra \}$. The definition of these states is closely related to
the way we could possibly measure them and to the way we prepare the system
in a certain state. That is why we shall start by saying how we can
initialize the qubit and measure it.

To initialize the qubit we put charge $1/2$ on antidot 1
\cite{sarma-freedman-nayak}. This can be done by
adding one quantum $h/e$ of magnetic flux in the vicinity of antidot 1 (e.g.,
by a solenoid piercing the antidot). A detailed analysis of the
tunneling situation in the stable strong-coupling regime in a Pfaffian
antidot with Aharonov--Bohm flux can be found in Ref.~\cite{5-2AB}.
The FQH liquid containing the antidot responds  by localizing a charged
excitation on the antidot border carrying one flux quantum. Because of the
fundamental FQH effect relation  between the magnetic flux and electric
charge\footnote{as usual we consider only the fractional part of the filling
	factor $\nu=2+1/2$ 	corresponding to the top-most Landau level}
\[
	Q_{\mathrm{el}} =\nu \Phi , \quad \nu=\frac{1}{2}
\]
 the charge of such an excitation is $1/2$. There are only two
 allowed charge $1/2$ excitations of the Pfaffian FQH state which could be
 localized on the antidot: in the notation of Refs.~\cite{5-2,5-2AB} these are
 represented by the field operators
 \beq \label{01-1/2}
	 \np{\e^{i\frac{1}{\sqrt{2}} \phi(z)} } \quad \mathrm{and} \quad
	 \psi(z) \np{\e^{i\frac{1}{\sqrt{2}} \phi(z)}}  ,  \quad \Phi=1, \quad Q_{\mathrm{el}}=1/2,
 \eeq
 where  $\psi(z)$ is the Ising model Majorana fermion and the vertex
 exponent of the normalized boson field $\phi(z)$ represents the u(1) part
 of the excitation.
Now we can envision the following computational basis
 \beqa \label{comp-basis}
|0\ra \quad &\longleftrightarrow& \quad \mathrm{if \ the \ charge \ } 1/2 \
\mathrm{state \ is}\quad \np{\e^{i\frac{1}{\sqrt{2}}\phi(z) } } \nn
|1\ra \quad &\longleftrightarrow& \quad \mathrm{if \ the \ charge} \ 1/2  \
\mathrm{state \ is}  \quad \psi(z)\np{\e^{i\frac{1}{\sqrt{2}}\phi(z) } },
 \eeqa
or, in other words, the state of the qubit is $|0\ra$ if the Majorana
fermion is not occupied and $|1\ra$ if it is occupied.
Note that the states corresponding to the fields in Eq.~(\ref{01-1/2})
cannot form coherent superpositions
because they belong to different superselection sectors inside the
Neveu--Schwarz sector (when the flux threading antidot 1 is one quantum
in the experimental  setup), which correspond to
their different fermion parity.

In Table~\ref{tab:sectors} we give for convenience the list of the
$6$ topologically inequivalent sectors (quasiparticles) for the Pfaffian FQH 
state. Note that the topological sectors in which the chiral fermion parity is 
well-defined contain quasiparticles with both positive and negative parities 
because they can be obtained from each other by adding an electron. The 
quasiparticle spectrum is obtained from the chiral partition functions of 
the corresponding topological sectors, see Refs.~\cite{5-2,5-2AB} for more 
details.
\begin{table}[htb]\label{tab:sectors}
\caption{Topologically inequivalent quasiparticles in the Pfaffian 
FQH state and their quantum numbers: electric charge $Q$,
neutral-sector chiral fermion parity $\gamma_F$,  
conformal dimension $\D$ and quantum statistics 
$\theta/\pi=2\Delta \ \mathrm{mod} \ 2$ }
\begin{tabular}{|l||c|c|c|c|c|}
\hline
\textbf{Particles in the same}  & \textbf{Fields} & \textbf{Charge} & 
\textbf{Parity} & \textbf{CFT dim.} & \textbf{Quant. stat.}  \\
 \textbf{Topological Sector} &  & $Q$ & $\gamma_F$ & $\Delta$&  $\theta/\pi$ \\
\hline\hline
vacuum & $\I$ & $0$ & $+$& $0$ & $0$ \\
hole/electron   & $ \e^{\pm i\sqrt{2}\phi}\psi$ & $\pm 1$ & $-$& $3/2$& $1$\\
\hline
 quasihole (vortex)&$ \e^{i\frac{1}{2\sqrt{2}}\phi}\sigma$  & $1/4$& undefined & $1/8$ & $1/4$ \\
\hline
quasiparticle &$ \e^{-i\frac{1}{2\sqrt{2}}\phi}\sigma$  & $\!\!\!\! -1/4$ & 
 undefined & $1/8$ & $1/4$ \\  
\hline
$+1$ flux &  $ \e^{i\frac{1}{\sqrt{2}}\phi}$ & $1/2$ & $+$& $1/4$ &$1/2$ \\
$-1$ flux $\times$ Majorana  & $ \e^{-i\frac{1}{\sqrt{2}}\phi}\psi$     & 
$\!\!\!\! -1/2$ &  $-$& $3/4$& $\!\!\!\! -1/2$\\
\hline
$-1$ flux  &  $ \e^{-i\frac{1}{\sqrt{2}}\phi}$ & $\!\!\!\! -1/2$ & $+$& $1/4$ &$1/2$ \\
$+1$ flux $\times$ Majorana   & $ \e^{i\frac{1}{\sqrt{2}}\phi}\psi$   & $1/2$ &
$-$ & $3/4$& $\!\!\!\! -1/2$\\
\hline
$+2$ flux  ($\kappa$-boson)&  $\e^{i\sqrt{2}\phi}$ & $1$ & $+$ &  $1$ & $0$ \\
Majorana fermion    &  $\psi$ & $0$ & $-$ &  $1/2$& 1\\
\hline
\end{tabular}
\end{table}
\subsection{Measurement of the qubit state}
\label{sec:measurement}
It was one of the  bright ideas  of Ref.~\cite{sarma-freedman-nayak} to use
the electronic Mach--Zehnder interferometry
\cite{Mach-Zehnder-Nature,bais-mach-zehnder,feldman-kitaev}
to determine  the state of the qubit.
More precisely, let us try to measure the diagonal conductance, $\sigma_{xx}$,
which is proportional to the probability for a charged particle to enter the
lower edge of the Hall bar in Fig.~\ref{fig:qubit} and to exit in out of the
upper  edge.
Because of the constrictions created by the front gates M and N, P and Q, there
are two alternative channels for a charged quasiparticle entering from the
lower edge to exit from the upper one: one is to tunnel between the front
gates M and N and  the other is to tunnel between P and Q. Therefore, to
leading order in the tunneling amplitudes $t_{MN}$ and $t_{PQ}$, which are
assumed to be very small, the diagonal conductance  would be proportional to
the interference of the two amplitudes \cite{sarma-freedman-nayak}.
Consider, for example the interference of the two
tunneling processes if the charge $1/2$ on antidot 1 does not contain
Majorana fermion. Then the amplitude for tunneling between P and Q must be
multiplied by the Aharonov--Bohm phase for the non-Abelian quasihole which
when tunneling between P and Q actually encircles the  charge $1/2$ state
 on antidot 1, i.e.,
\beq \label{AB}
\e^{2\pi i \Phi \, Q_{\mathrm{el}} } = \e^{i\frac{\pi}{2}} =i,
\eeq
because a quasiparticle with charge $Q_{\mathrm{el}}=1/4$ encircles
magnetic flux $\Phi=1$. Therefore, the diagonal conductance, which is proportional to
the modulus-square of the amplitude, reads
\beq \label{sigma_0}
	\sigma^{|0\ra}_{xx}\propto |t_{MN}+i\, t_{PQ}|^2 .
\eeq
If instead, the state of the qubit is $|1\ra$, i.e., there is a Majorana
fermion on antidot 1, then in addition to the Aharonov--Bohm phase (\ref{AB})
there would be a minus sign coming from the fact that the Ising model $\sigma$
filed is transported around the Majorana fermion. To see, in the CFT language,
why this minus sign appears,
we consider the operator-product-expansion
\[
\sigma(z)\psi(0)  \mathop{\sim}_{z \to 0} \frac{\sigma(0)}{\sqrt{2z}}\quad
\mathrm{so \ that}\quad
 \sigma(z)\psi(0)  \to
-\sigma(z)\psi(0) \quad \mathrm{when}\quad  z \to \e^{2\pi i}z  ,
\]
i.e., when $\s$ is transported adiabatically around $\psi$.
Thus, the diagonal conductance measurement in the state $|1\ra$ gives
\beq \label{sigma_1}
	\sigma^{|1\ra}_{xx}\propto |t_{MN}-i \, t_{PQ}|^2 .
\eeq
Note that the two different interference patterns, Eqs.~(\ref{sigma_0}) and
(\ref{sigma_1}) of the diagonal conductance are very well distinguished
experimentally due to the high visibility of the Mach--Zehnder interferometry
\cite{Mach-Zehnder-Nature}.
\subsection{Splitting the $1/2$ charge: finalizing the qubit}
\label{sec:splitting}
Despite that we can efficiently measure the two states in the computational
basis this is still not sufficient for quantum computation. The reason is
that we need to form coherent superpositions of the states $|0\ra$ and  $|1\ra$
which is not allowed for the charge $1/2$ states (\ref{comp-basis}) due to the
fermion parity superselection rule. In order to circumvent this difficulty
Das Sarma et al. have made another interesting proposal
\cite{sarma-freedman-nayak}:
 to split the charge $1/2$ state into $1/4 \times 1/4$
state by transferring one charge $1/4$ from antidot 1 to antidot 2.
This is indeed possible, if one applies voltage between the two antidots,
 because the most relevant quasiparticle for tunneling through the bulk of
 the Pfaffian FQH liquid carries charge $1/4$. This quasiparticle is
 non-Abelian and contains a $\sigma$ field from the Ising model because
 there is no other charge $1/4$ quasiparticle and the non-Abelian one has
 the minimal CFT dimension \cite{mr,nayak-wilczek,5-2}.
 Now the state on antidots 1 and 2 is equivalent
 to $\sigma(\eta_1)\sigma(\eta_2)$ which together with the quasiholes at
 $\eta_3$ and $\eta_4$ correspond to a 4-quasihole wave function, which belongs
 to a two-dimensional space as discussed in the beginning of
Sect.~\ref{sec:sarma}.

While the charge $1/2$ states (\ref{comp-basis}) cannot form coherent
superpositions, because they belong to different superselection sectors,
for the 4 quasiholes configurations we can consider the superposition of
the states  obtained by  splitting of the vacuum in two different ways,
namely
\beqa
\I \to \I \times \I \to (\s\times\s) \times (\s\times\s)\nn
\I \to \psi \times \psi \to (\s\times\s) \times (\s\times\s),
\eeqa
because the two 4-quasihole states now belong to the same superselection
sectors.

One natural question arises in this charge splitting procedure:
\textit{does the state of  the qubit remain the same in the process of
transferring
one charge $1/4$   from antidot 1 to antidot 2 or it changes?}
Or, even,  more philosophically: \textit{does the system with 4 $\sigma$ fields
has something in common  with the computational basis of our charge $1/2$
state plus 2 additional	$\sigma$ fields, or it is completely different?}

The original idea of Ref.~\cite{sarma-freedman-nayak} was that the state of
the qubit does not change during the splitting procedure because the pairs
of quasiholes in the Pfaffian state share a pair of Majorana fermions zero
modes, $(\psi_0, \bar{\psi}_0)$,
whose combined (left- plus right- moving) fermion parity is supposed to be
conserved. That's why the read-out should give the same results as before
splitting unless some Majorana fermions could tunnel from the edge or from
another antidot,
however, these processes are exponentially suppressed because the quasiholes are
supposed to be well separated and far from the edges.

In Sect.~\ref{sec:read-out} we shall explicitly derive the readout results for the
4-quasiholes wave functions and will find conditions under which this is in
agreement with Ref.~\cite{sarma-freedman-nayak}.
\subsection{Four-quasiholes wave functions}
\label{sec:4-qh}
As we shall see in Sect.~\ref{sec:read-out}, the topological phase, which
determines the conductance interference pattern, can be obtained from the
monodromy matrices appearing in the adiabatic transport of some quasiholes
around others, so that we need the wave functions explicitly.
The wave function  for even number  $N$ of holes (or electrons) at positions
$z_1, \ldots, z_N$ containing 4 quasiholes
at positions $\eta_1,\ldots, \eta_4$, can be realized as a correlation function
\[
\Psi_{4\qh}(\eta_1,\eta_2,\eta_3,\eta_4; \{z_i\})=
\la \psi_{\mathrm{qh}}(\eta_1)\psi_{\mathrm{qh}}(\eta_2)\psi_{\mathrm{qh}}(\eta_3)
\psi_{\mathrm{qh}}(\eta_4)\prod_{i=1}^N
\psi_{\mathrm{hole}}(z_i) \ra
\]
 of the field operators corresponding to creation of holes and quasiholes
\beq \label{fields}
\psi_{\mathrm{hole}}(z)= \psi(z) \, \np{\e^{i \sqrt{2}\phi(z)}  } \quad
\mathrm{and}\quad
\psi_{\mathrm{qh}}(\eta)= \s(\eta)\, \np{\e^{i\frac{1}{2\sqrt{2}} \phi(\eta)}},
\eeq
respectively, where $\sigma(\eta)$ is the chiral spin field in the Ising model
of dimension $1/16$ and  $\psi(z)$ is the right-moving Majorana fermion
 in the chiral Ising model.
 It can be expressed as
\beqa \label{4qh}
\Psi_{4\mathrm{qh}}(\eta_1,\eta_2,\eta_3,\eta_4; z_1, \ldots, z_N)=
\left\la \sigma(\eta_1)\sigma(\eta_2)\sigma(\eta_3)\sigma(\eta_4)
\prod_{j=1}^N \psi(z_j)\right\ra \times \nn
\prod_{1\leq a<b\leq 4} \eta_{ab}^\frac{1}{8}  \
 \prod_{i=1}^{N} \sqrt{(z_i-\eta_1)(z_i-\eta_2) (z_i-\eta_3)(z_i-\eta_4)}
\prod_{1\leq i<j\leq N} z_{ij}^2,
\eeqa
where the average sign $\la \cdots\ra$ now represents the vacuum expectation
value in the chiral Ising model and the last three product factors in
Eq.~(\ref{4qh})
come from the $u(1)$ components of the holes $\np{\exp(i\sqrt{2}\phi(z))}$ and
of the quasiholes $\np{\exp(i\phi(z)/2\sqrt{2})}$.  We used here the notation
$\eta_{ab}\equiv \eta_{a} -\eta_{b}$ for $a\neq b $ and $z_{ij}=z_i-z_j$ for
$i\neq j$.

One important detail is that the chiral field $\s(\eta)$ does not have a
definite fermion parity because of the non-Abelian fusion rule
(\ref{fusion}) which mixes states with different fermion parities.
Nevertheless it would be convenient to introduce two chiral fields
$\s_{\pm}$ of dimension $\Delta_\pm=1/16$ and opposite parities \cite{5-2}
\beq \label{gamma_F2}
\gamma_F \sigma_{\pm}(\eta) \gamma_F = \pm  \sigma_{\pm}(\eta) , \quad
\eeq
in terms of which the non-Abelian  fusion rule (\ref{fusion}) necessarily
splits into two Abelian channels
\beq \label{sigma_pm}
\s_\pm \times \s_\pm =\I, \quad \s_\pm \times \s_\mp=\psi.
\eeq
The $\s$ filed entering Eq.~(\ref{4qh}) is then identified with
\beq\label{sigma}
\s(\eta)=\frac{\s_+(\eta) +\s_-(\eta)}{\sqrt{2}}.
\eeq
In order to obtain the 4-qh wave function (\ref{4qh})
we carefully repeat the arguments of Ref.~\cite{nayak-wilczek}
in which the chiral many-body wave functions of the Pfaffian state
are obtained by bosonization techniques from the $c=1$ complex Ising model.
The  final result is
\beq \label{true}
\Psi_{4\mathrm{qh}}(\eta_1,\eta_2,\eta_3,\eta_4; z_1, \ldots, z_N)
=\Psi_{4\mathrm{qh}}^{(0)}+\Psi_{4\mathrm{qh}}^{(1)},
\eeq
where we have used the notation of
Refs.~\cite{sarma-freedman-nayak,nayak-wilczek}
($\Psi_{4\mathrm{qh}}^{(0)}\equiv |0\ra$, $\Psi_{4\mathrm{qh}}^{(1)}\equiv |1\ra$
 would be our computational basis in the 4-qh wave function's space)
\beq\label{01}
\Psi_{4\mathrm{qh}}^{(0,1)}=
\frac{\left(\eta_{13}\eta_{24}\right)^{\frac{1}{4}}}{\sqrt{1\pm \sqrt{x}}}
\left(\Psi_{(13)(24)}\pm \sqrt{x}\,  \Psi_{(14)(23)} \right)
\eeq
 with $x$ being a CFT invariant crossratio \cite{CFT-book}
\beq \label{x}
x=\frac{\eta_{14}\eta_{23}}{\eta_{13}\eta_{24}} \quad \mathrm{and} \quad
\left( \eta_{ab}=\eta_a-\eta_b \right),
\eeq
\beqa
\Psi_{(ab)(cd)}&=&\Pf\left(
\frac{(z_i-\eta_a)(z_i-\eta_b)(z_j-\eta_c)(z_j-\eta_d)+
(i\leftrightarrow j)}{z_i-z_j}\right) \times \nn
&\times& \prod_{1\leq i<j\leq N} (z_i - z_j)^2 ,   \nonumber
\eeqa
where $\{a,b,c,d\}$ is a permutation of $\{1,2,3,4\}$ satisfying
$a<b$ and $c<d$. The Pfaffian of an anti-symmetric matrix $M_{ij}$
of even dimension $N$ is defined as
\[
\Pf \left( M_{ij}\right)= \frac{1}{2^{N/2}((N/2)!)^2} \sum_{\s\in S_N}
\mathrm{sign}(\s) \prod_{k=1}^{N/2} M_{\s(2k-1)\s(2k)}.
\]
It is worth stressing that the space of the 4-quasihole wave functions for
fixed positions of the quasiholes is two-dimensional.
The second independent wave function could be obtained from
the first one, Eq.~({\ref{true}}), by transporting $\eta_3$ around
 $\eta_4$ (i.e., $\eta_{34}\to \e^{2\pi i} \eta_{34}$)
 which transforms $\Psi_{4\mathrm{qh}}^{(0)}\to \Psi_{4\mathrm{qh}}^{(0)}$ and
 $\Psi_{4\mathrm{qh}}^{(1)}\to -\Psi_{4\mathrm{qh}}^{(1)}$ so that
\beq \label{true2}
\widetilde{\Psi}_{4\mathrm{qh}}(\eta_1,\eta_2,\eta_3,\eta_4; z_1, \ldots, z_N)
=\Psi_{4\mathrm{qh}}^{(0)}-\Psi_{4\mathrm{qh}}^{(1)}.
\eeq
It appears, however, that in some situations, such as the qubit initialization of
Ref.~\cite{sarma-freedman-nayak}, the 4-qh wave function may be driven
directly in the state $\Psi_{4\mathrm{qh}}^{(0)}$ or
$\Psi_{4\mathrm{qh}}^{(1)}$.
\section{Read-out in the TQC scheme of Das Sarma et al: the measurement of the
	4-quasiholes qubit state}
\label{sec:read-out}
As we have seen in Sect.~(\ref{sec:measurement}) the read-out
of the qubit is performed by interference measurement of the
diagonal conductance, for which we need the monodromies of the corresponding
wave functions.
In this section we shall compute the explicit monodromies of the 4-qh wave
functions (\ref{01}) as well as for (\ref{true}) and (\ref{true2}). Then,
in Sect.~\ref{sec:1-2-read-out}, we
shall compute the corresponding monodromies for the two charge $1/2$ states,
which can be obtained by fusing quasiholes with coordinates $\eta_1$ and
$\eta_2$.

\begin{figure}[htb]
\centering
\includegraphics*[bb=20 390 595 730,width=\textwidth]{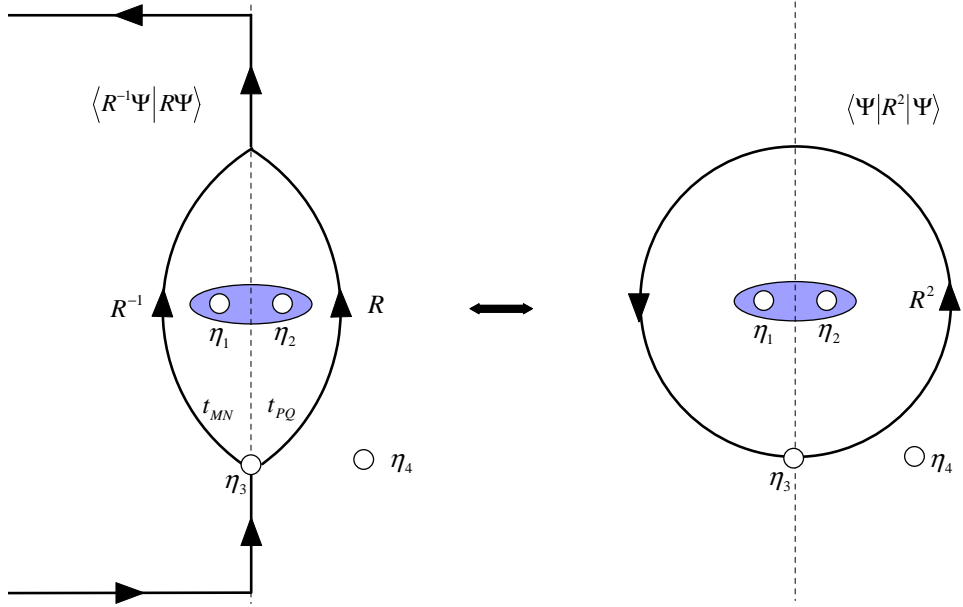}
 \caption{Conductance interference term in the read-out procedure expressed by
 	the braids $R$ (corresponding to tunneling through P and Q), $R^{-1}$
	(corresponding to tunneling through M and N) and the monodromy $R^2$.}
	\label{fig:read-out}
\end{figure}

Assuming that our qubit is formed by the quasiholes with coordinates
$\eta_1$ and $\eta_2$ we can interpret the quasihole with coordinate $\eta_3$
tunneling either through M and N or through P and Q
as generating the interference pattern in the longitudinal conductance,
like in the non-Abelian
Mach--Zehnder interferometer \cite{bais-mach-zehnder}.  In more detail,
if the third quasihole tunnels through M and N, as shown on
Fig.~\ref{fig:read-out},
 this could be interpreted as a
clockwise braid of this quasihole with the qubit, while if it tunnels through
P and Q this gives rise to a counter-clockwise braid (denoted below as $R$; cf.
Ref.~\cite{bais-mach-zehnder}) so that the quantum
amplitude for these two processes is (we have absorbed all dynamical phase
factors for the two paths in the corresponding  amplitudes $t_{MN}$ and
$t_{PQ}$)
\[
|A \ra =t_{MN} \, R^{-1} |(\eta_1,\eta_2),\eta_3, \eta_4 \ra +
t_{PQ}\, R |(\eta_1,\eta_2),\eta_3, \eta_4 \ra
\]
and the longitudinal conductance would be
\[
\sigma_{xx}\propto \la A| A \ra  = |t_{MN}|^2 +  |t_{PQ}|^2 +
2\Re \left( t^{*}_{MN}t_{PQ}   \la \Psi| R^2 |\Psi\ra \right),
\]
where $|\Psi\ra=| (\eta_1,\eta_2),\eta_3, \eta_4 \ra$ is the 4-qh state in
which
the qubit  is formed by the quasiholes with positions $\eta_1$ and $\eta_2$,
and  we have used  the unitarity of the braid operator $R$ and
$R^2$ is the corresponding  monodromy operator. The operator $R^2$ is actually
the operator which
\textit{takes the quasihole with coordinate  $\eta_3$ around  those with
$\eta_1$ and $\eta_2$ in counter-clockwise direction}.
Note that if we fuse the quasiholes forming the qubit, i.e.,
 $\eta_1\to \eta_2$, and the state is $\I$
then the matrix element is $+i$, while if they fuse to $\psi$ the matrix
element is $-i$, which reproduce the interference results
in Sect.~\ref{sec:measurement}.
It is worth stressing that there is a remarkable relation between
the expectation value of the monodromy operator $R^2$, corresponding to the
adiabatic transport of particle with label $a$ around particle with label $b$,
 and the modular $S$ matrix \cite{CFT-book} associated with any
 rational CFT \cite{bonderson-12-5}:
\[
 \la \Psi| R^2 |\Psi\ra  =\frac{S_{ab}S_{00}}{S_{0a}S_{0b}},
\]
where the label $0$ corresponds to the vacuum.
Working with modular $S$ matrices is very convenient because they are
explicitly known for almost all rational CFT related to the FQH effect.

Thus we see that the read-out or measurement procedure for the 4-qh wave
functions consists in taking the quasihole
with coordinate $\eta_3$ around the qubit
\cite{sarma-freedman-nayak,nayak-private},
i.e., around the two quasiholes with coordinates  $\eta_1$ and $\eta_2$.
\footnote{Alternatively,
as a matter of choice, 	one could use the quasihole
with the  coordinate $\eta_4$ to encircle the quasiholes localized
on antidots $1$ and $2$, which would lead to the same results, however,
we will stick  here to the notation of Ref.~\cite{sarma-freedman-nayak}).}

\begin{figure}[htb]
\centering
\includegraphics*[bb=0 340 540 500,width=\textwidth]{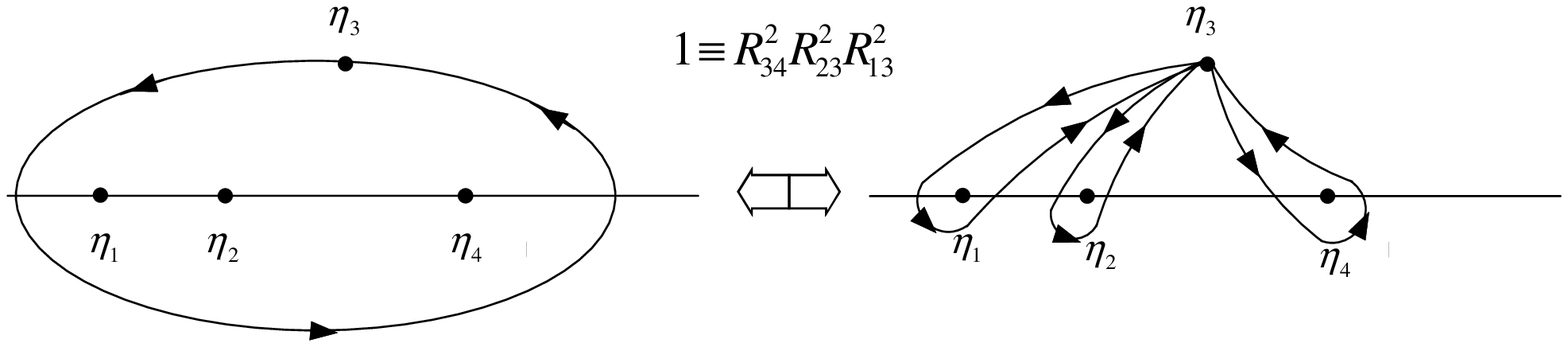}
 \caption{The trivial monodromy of the 4-quasihole wave function, when one
 	quasihole is transported around all others,  allows to compute the read-out
	as taking $\eta_3$ around $\eta_4$.}
	\label{fig:monodromy}
\end{figure}

It is worth stressing that the 4-qh wave function basis (\ref{01}) is very 
convenient from the point of view of the adiabatic transport because the 
functions $\Psi_{(ab)(cd)}$ have no Berry phases \cite{nayak-wilczek} so that 
the transport effects could be explicitly determined from the
monodromies of the multivalued function
\[
f(z)=  \sqrt{1 \pm \sqrt{z}}, \quad z\in \C ,
\]
entering the denominator in Eq.~(\ref{01}).
Notice, however,  that ``taking $\eta_3$ around $\eta_1$ and $\eta_2$'' in the
functions (\ref{f}) for $z=x$ with $x$ defined in Eq.~(\ref{x}), which are
simply proportional to the 4-point functions of the chiral spin field in the
Ising model, is equivalent to the adiabatic transport of $\eta_3$ around
$\eta_4$ alone. This is because the monodromy for the transport of $\eta_3$
around $\eta_{1}$, $\eta_{2}$ and $\eta_{3}$ is trivial, as shown on
Fig.~\ref{fig:monodromy},
since the contour on the left can be contracted to a point
(at infinity) without passing through any other singularity.
Therefore it would be simpler to compute  the read-out transformation of
the  functions (\ref{f}) for $z=x$ by
\beq \label{read-out}
R_{34}^2:  \qquad  \eta_{34} \to
\eta'_{34}= \lim_{t\to 1_-}  \e^{ 2\pi i\, t}\eta_{34}, \quad t \in [0, 1). \quad
\eeq
The transformation (\ref{read-out}) obviously preserves the absolute
value of the crossratio
\beq \label{tx}
		\tx \equiv \frac{\eta_{12}\eta_{34}}{\eta_{13}\eta_{24}} = 1-x
		\quad \Leftrightarrow \quad x=1-\tx,
\eeq
where $x$ is defined in Eq.~(\ref{x}), because all other 
$\eta'_{ab}=\eta_{ab}$, i.e., the contour for transportation
of $\tx$, corresponding to Eq.~(\ref{read-out}), is a circle with center
at $\tx=0$ and radius $|\tx|$. Therefore the result crucially depends on
whether $|\tx|$ is bigger or smaller than 1.
Thus we need  to consider the behavior of the functions
\beq \label{f}
	f_\pm(\tx)=\sqrt{1\pm \sqrt{1-\tx}}
\eeq
under the transformation (\ref{read-out}). Yet, it is more convenient
to first analyze Eq.~(\ref{f}) as a function of $z$ and then just outline what
happens when we change  $z=1-\tx$.

The multivalued function (\ref{f}) has two separate branching points:
one is $z=0$ which, if encircled by the continuation contour,
would change the sign of the inner root, as can be seen from Eq.~(\ref{z-0})
in Appendix~\ref{app:Laurent}, i.e., 
\[
\sqrt{1\pm \sqrt{z}} \ \ \to  \ \ \sqrt{1\mp \sqrt{z}}, \quad 
\mathrm{for} \quad 
z\to \e^{2\pi i} z \quad \mathrm{and} \quad |z|<1 .
\]
However, the second function,
corresponding to the minus sign under the square root in Eq.~(\ref{f}), 
has one more branching point at $z=1$ which, if encircled
in the process of the analytic continuation changes the sign of the outer root,
i.e., 
\[
\sqrt{1\pm \sqrt{z}} \ \  \rightarrow  \ \ \pm \sqrt{1\pm \sqrt{z}}, 
\quad \mathrm{for} \quad 
(z-1)\to \e^{2\pi i} (z-1) \ \ \mathrm{and} \ |z-1|<1 .
\]
This can be easily seen from the Laurent-mode expansion for $|z-1|< 1$, by
looking at Eq.~(\ref{z-1}) in Appendix~\ref{app:Laurent}.

Now consider what happens when we change $z=1-\tx$.
When $\tx\to \e^{2\pi i t}\tx$, and $t$ goes from 0 to 1, this transports
$\tx$ along a circle of radius $|\tx|$ and center at $\tx=0$.
Then $-\tx$ is transported along the same  circle,
though with a phase shift of $\pi$. Thus $z=1-\tx$ is transported along
a circle of the same radius $|\tx|$, however translated to be centered at
$z=1$ as shown on Figures~\ref{fig:smaller} and \ref{fig:bigger}.

Therefore  we need to consider both cases, $|\tx|<1$ and   $|\tx|>1$,
separately  in more detail. Of course, one may consider other contours
which are homotopic
to that of Eq.~(\ref{read-out}) but do not preserve the absolute value of
$\tx$. In that case the results would be the same as those obtained from
Eq.~(\ref{read-out}), only the conditions on  $|\tx|$ must be replaced by
the homotopic condition  whether these contours encircle both
branching points $z=0$ and $z=1$ or only one of them.
\subsection{The read-out for $|\tx|<1$: take $\eta_3$ around $\eta_4$  }
\label{sec:read-out-34-1}
\begin{figure}[htb]
\centering
\includegraphics*[bb=50 430 480 680,width=12cm]{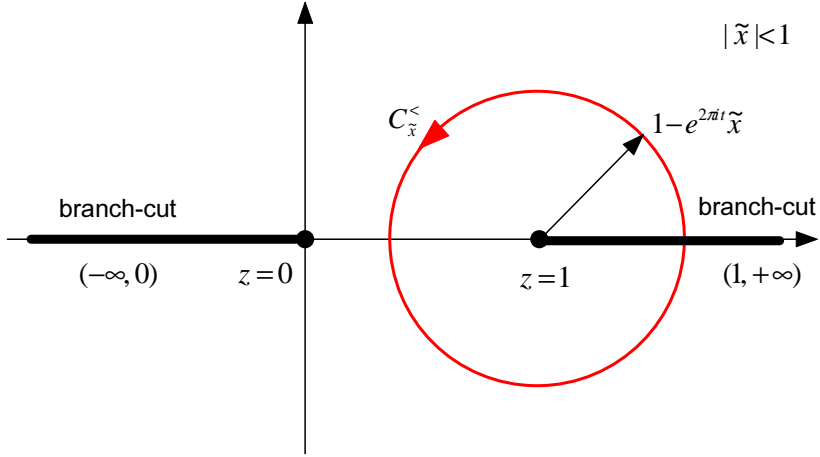}
 \caption{The contour $C^{<}_{\tx}$ used for the analytic continuation
of Eq.~(\ref{f}) when $|\tx|<1$. As it encloses only the branching point
at $z=1$, going along this contour only changes the outer-root sign of
$f_-(z)$.}
\label{fig:smaller}
\end{figure}

According to our analysis, when  $|\tx|<1$,  the read-out transformation
(\ref{read-out}) corresponds to transporting  the crossratio (\ref{tx}) along a
contour which encircles only the branching point at $z=1$,
\beq \label{smaller}
C_{\tx}^{<} = \left\{ z=1-\e^{2\pi i\, t} \tx \ | \  0 \leq t \leq 1,
\quad |\tx|<1 \right\} ,
\eeq
as shown in Fig.~\ref{fig:smaller}.
We have introduced two branch-cuts for the function (\ref{f})
like in Fig.~\ref{fig:smaller}.
Because the 4-quasihole wave functions $\Psi_{(13)(24)}$ and $\Psi_{(14)(23)}$
have no branch-cuts in $\eta_{ab}$, they acquire no phase
under the transformation (\ref{read-out}), and the only phases come from the
prefactor in Eq.~(\ref{01}) containing $\eta_{ab}$ and $x$.

Since $|\tx|<1$ and the transformation (\ref{read-out})
transports $\tx$ along the contour (\ref{smaller}),
which is shown on Fig.~\ref{fig:smaller}, this transformation only changes
the outer-roots signs, i.e.,
\[
R_{34}^2  \left(\matrix{\sqrt{1 +\sqrt{1-\tx}} \cr
\sqrt{1 -\sqrt{1-\tx}}  } \right) =
\left(\matrix{1 & \ \ 0 \cr  0& -1} \right)
\left(\matrix{\sqrt{1 + \sqrt{1-\tx}} \cr  \sqrt{1 -\sqrt{1-\tx}}  } \right) .
\]
 In order to find the action of the transformation (\ref{read-out})
in the  basis
$\left \{ \Psi_{4\mathrm{qh}}^{(0)}, \Psi_{4\mathrm{qh}}^{(1)} \right\}$
we only have to add the phase coming from $\eta_{13}^{1/4}$ in Eq.~(\ref{01})
which is $\e^{i\pi/2}$. Notice that we use ``taking $\eta_3$ around $\eta_4$''
as the readout prescription only for the functions (\ref{f}) because these 
are the functions for which the total monodromy is trivial. For all other 
functions, including fractional powers of $\eta_{ab}$, we still use as the 
read-out ``taking $\eta_3$ around $\eta_1$ and 
$\eta_2$''. Thus we obtain the read-out transformation as
\beq \label{U_<}
  U^{(|\tx|<1)}_{\mathrm{read-out}}
	\left(\matrix{ \Psi_{4\mathrm{qh}}^{(0)}  \cr
	\Psi_{4\mathrm{qh}}^{(1)} }\right) =
	\left(\matrix{i & \ \ \ 0 \cr  0& -i} \right)
	\left(\matrix{ \Psi_{4\mathrm{qh}}^{(0)}  \cr
	\Psi_{4\mathrm{qh}}^{(1)} }\right)
	\quad \mathrm{for} \quad
	\left|\frac{\eta_{12}\eta_{34}}{\eta_{13}\eta_{24}}\right| <1  .
\eeq
Notice that this ``$\pm i$'', on the diagonal of the  matrix  in
Eq.~(\ref{U_<}),
is exactly the topological phase that appears
for the quasiparticles traveling along the contour passing through the front
contacts P and Q, see  Fig.~\ref{fig:qubit}, in the  diagonal conductance
measurement   so that
\beq \label{sigma_xx1}
\sigma_{xx}^{|\tx|<1}\propto |t_{MN} \pm i \, t_{PQ}|^2 , \quad
\mathrm{with} \ "+" \ \mathrm{for} \ |0\ra \ \mathrm{and} \ \ ``-" \
\mathrm{for} \ |1\ra.
\eeq
\subsection{The case $|\tx|>1$ }
\label{sec:read-out-34-2}
\begin{figure}[htb]
\centering
\includegraphics*[bb=60 380 490 730,clip,width=12cm]{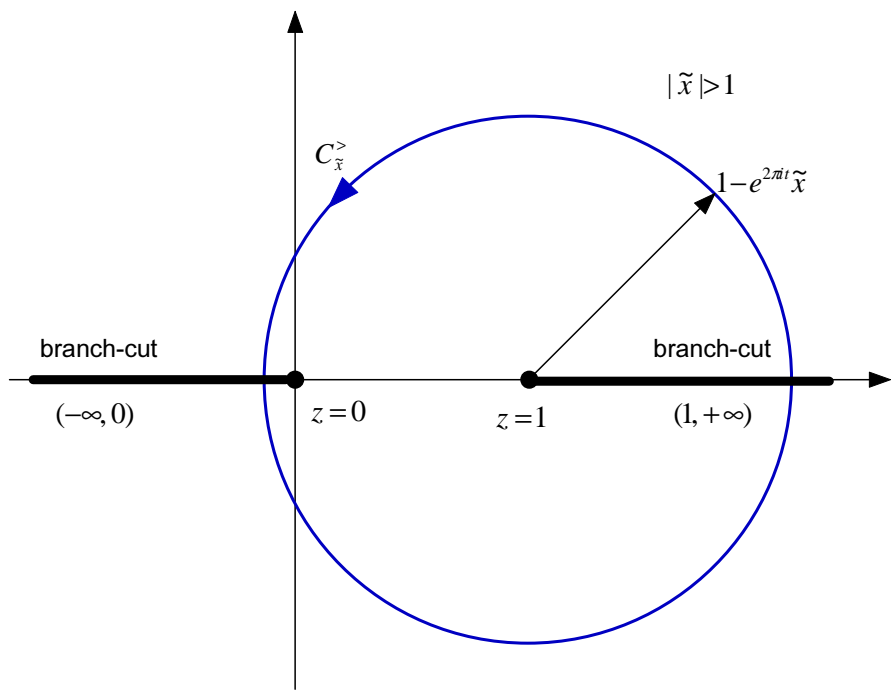}
 \caption{The contour $C^{>}_{\tx}$ used for the analytic continuation
of Eq.~(\ref{f}) when $|\tx|>1$. Because it encloses both branching point
at $z=1$ and $z=0$ going along this contour changes the outer-root sign
whenever the inner one is ``$-$''  and flips the inner-root sign of $f_\pm(z)$.}
\label{fig:bigger}
\end{figure}

Alternatively,  if $|\tx|>1$, then the read-out transformation
(\ref{read-out}) leads to transporting $\tx$ along the following contour
\beq \label{bigger}
C_{\tx}^{>} = \left\{ z=1-\e^{2\pi i\, t} \tx \ | \  0 \leq t \leq 1,
		\quad |\tx|>1 \right\},
\eeq
which encircles both branching points at $z=1$ and $z=0$ as shown on
Fig.~\ref{fig:bigger}.
Therefore, the read-out transformations (\ref{read-out}) would change
the sign of the outer root whenever the inner sign
is ``$-$''  \textit{as well as would flip the  inner-roots signs}, i.e.,
\[
R_{34}^2
\left(\matrix{\sqrt{1 +\sqrt{1-\tx}} \cr \sqrt{1 -\sqrt{1-\tx}} }\right) =
\left(\matrix{0 &  1 \cr  -1& 0} \right)
\left(\matrix{\sqrt{1 + \sqrt{1-\tx}} \cr  \sqrt{1 -\sqrt{1-\tx}}  } \right) .
\]
Adding again  the Abelian phase $\e^{i\pi/2}$, coming from $\eta_{13}^{1/4}$ in
Eq.~(\ref{01}) we obtain the read-out transformation as
\beq \label{U_>}
  U^{(|\tx|>1)}_{\mathrm{read-out}}
	\left(\matrix{ \Psi_{4\mathrm{qh}}^{(0)}  \cr
	\Psi_{4\mathrm{qh}}^{(1)} }\right) =
	\left(\matrix{0 & i \cr -i & 0} \right)
	\left(\matrix{ \Psi_{4\mathrm{qh}}^{(0)}  \cr
	\Psi_{4\mathrm{qh}}^{(1)} }\right)
	\quad \mathrm{for} \quad
	\left|\frac{\eta_{12}\eta_{34}}{\eta_{13}\eta_{24}}\right| > 1  .
\eeq
Now  the  diagonal conductance measurement  might  be different
from Eq.~(\ref{sigma_xx1}), as it crucially depends on the overlap between
the states (\ref{01}). If the two states $\Psi_{4\mathrm{qh}}^{(0,1)}$
are orthogonal, which is a fundamental requirement in quantum theory, then the
monodromy average
$\la \Psi_{4\mathrm{qh}}^{(0,1)} | R^2_{34} |\Psi_{4\mathrm{qh}}^{(0,1)}\ra$,
determining the interference pattern, would
vanish and the diagonal conductance
could not distinguish between the two states in the computational basis.
Thus we see that the Mach--Zehnder interference measurement could only work  for
 $\left|\eta_{12}\eta_{34}\right| < \left| \eta_{13}\eta_{24}\right|$.

Using the CFT invariance of $\tx$, one can prove that the
absolute value of the crossratio $\tx$ depends on the absolute values
of the quasiholes positions, i.e.,
\[
|\tx| <1  \quad \Leftrightarrow \quad |\eta_{3}|<|\eta_{2}|, \quad
\mathrm{while}  \quad
|\tx| > 1  \quad \Leftrightarrow \quad |\eta_{3}|>|\eta_{2}|.
\]
\begin{rem}
The main result in this Section has very important implications for the experimental 
realization of the topological quantum computer in FQH systems at $\nu=5/2$.
The read-out procedure crucially depends on the absolute value of the crossratio
(\ref{tx}). When $|\tx| <1$ the TQC scheme of Ref.~\cite{sarma-freedman-nayak} is 
going to work as originally proposed, while if $|\tx| > 1$ this scheme would fail.
As the absolute value of $\tx$ depends on the quasiholes positions, this certainly 
gives some hints on how the antidots should be arranged in order for the read-out 
procedure to work  as in Ref.~\cite{sarma-freedman-nayak} and,  as we shall see later,
this is also related to whether the NOT gate could be executed in the Pfaffian qubit 
or not.
\end{rem}
\subsection{Orthogonality of the 4-quasiholes wave function
	$\Psi_{4\mathrm{qh}}^{(0,1)}$ }
\label{sec:orthogonality}
One of the fundamental requirement in quantum theory
is that the states $|0\ra$, $|1\ra$ are not only linearly independent but
in fact orthogonal. Otherwise there exists
no measurement that can reliably distinguish these states \cite{nielsen-chuang}.
Here we will  demonstrate that the   4-quasiholes states (\ref{01}) are
indeed orthogonal, at least  for $|\tx|<1$.
This seems natural because, as we already know the  Mach--Zehnder
interferometer can distinguish the states (\ref{01}), hence
these states should be orthogonal. However, it is still worth verifying
this directly if possible.

The point is that  for $|\tx|<1$ we have, according to Eq.~(\ref{U_<}),
$U\Psi_{4\mathrm{qh}}^{(0)} =i \, \Psi_{4\mathrm{qh}}^{(0)}$ and
$U\Psi_{4\mathrm{qh}}^{(1)} =-i \, \Psi_{4\mathrm{qh}}^{(1)}$, hence,
$U^\dagger\Psi_{4\mathrm{qh}}^{(0)} =-i \, \Psi_{4\mathrm{qh}}^{(0)}$ and
$U^\dagger\Psi_{4\mathrm{qh}}^{(1)} =i \, \Psi_{4\mathrm{qh}}^{(1)}$, where
$U\equiv U_{\mathrm{read-out}}$. In order to find the overlap
$\left( \Psi_{4\mathrm{qh}}^{(1)},  \Psi_{4\mathrm{qh}}^{(0)}\right)$  we first plug
inside it the operator $U$ and use the above relations, i.e.,
on one side we have
\beq \label{orth-1}
\left( \Psi_{4\mathrm{qh}}^{(1)}, U \Psi_{4\mathrm{qh}}^{(0)}\right) =
i \, \left( \Psi_{4\mathrm{qh}}^{(1)}, \Psi_{4\mathrm{qh}}^{(0)}\right)
\eeq
while, on the other side, using the definition of the Hermitean conjugation and
the anti-linearity of the inner product with respect to its first argument,
it is equal to
\beq \label{orth-2}
\left( U^\dagger \Psi_{4\mathrm{qh}}^{(1)},  \Psi_{4\mathrm{qh}}^{(0)}\right)  =
\left( i \, \Psi_{4\mathrm{qh}}^{(1)},  \Psi_{4\mathrm{qh}}^{(0)}\right)  =
-i \, \left( \Psi_{4\mathrm{qh}}^{(1)},  \Psi_{4\mathrm{qh}}^{(0)}\right).
\eeq
Obviously the right-hand sides of Eqs.~(\ref{orth-1}) and (\ref{orth-2}) 
could only be equal if
\[
\left( \Psi_{4\mathrm{qh}}^{(1)},  \Psi_{4\mathrm{qh}}^{(0)}\right)=0.
\]
\section{Measurement of the charge $1/2$ state's wave function}
\label{sec:1-2-read-out}
As we have seen in Sect.~\ref{sec:measurement}
the state of the qubit before splitting $1/2\to 1/4 \times 1/4$ is
$|0\ra$ if the Majorana fermion is absent or $|1\ra$ if  it is present
in the charge-$1/2$ state.
Using the fusion rules for two quasiholes  (\ref{fields})
\beq \label{qh-OPE}
\psi_\qh(\eta_1) \psi_\qh(\eta_2) \mathop{\simeq}_{\eta_{1}\to\eta_2}
\left(1+\sqrt{\frac{\eta_{12}}{2}} \psi(\eta_2)\right)
\e^{i\frac{1}{\sqrt{2}}\phi(\eta_2)}
\eeq
as well as the general fusion rule of vertex exponents
\[
\e^{i \lambda_a \phi(\eta_a)}  \e^{i \lambda_b \phi(\eta_b)}
\mathop{\simeq}_{\eta_a\to\eta_b}\eta_{ab}^{\lambda_a \lambda_b }
\e^{i \left(\lambda_a + \lambda_b\right)\phi(\eta_b) }
\]
we can deduce the form of the charge $1/2$ wave functions from the
4-quasihole wave function (\ref{4qh}) by taking $\eta_1\to\eta_2$.
Let us denote by $\Psi_0$  and  $\Psi_1$ the result of fusing the
two  quasiholes $\psi_\qh$ to $\e^{i\frac{\phi}{\sqrt{2}}}$ and
$\psi \e^{i\frac{\phi}{\sqrt{2}}}$ for $\eta_1\to\eta_2$ as in
Eq.~(\ref{qh-OPE}), respectively, i.e.,
\beqa \label{01-1-2}
\Psi_0&=&\left \la \s (\eta_3)\s(\eta_4)\prod_{i=1}^N \psi(z_i) \right \ra
\left \la \e^{i\frac{1}{\sqrt{2}}\phi(\eta_2)}
\e^{i\frac{1}{2\sqrt{2}}\phi(\eta_3)}
\e^{i\frac{1}{2\sqrt{2}}\phi(\eta_4)}
\prod_{i=1}^N \e^{i\sqrt{2}\phi(z_i)} \right \ra \nn
\Psi_1&=&\sqrt{\frac{\eta_{12}}{2}}
\left \la  \psi(\eta_2)\s(\eta_3)\s(\eta_4)\prod_{i=1}^N \psi(z_i) \right \ra
\left \la \e^{i\frac{\phi(\eta_2)}{\sqrt{2}}}
\e^{i\frac{\phi(\eta_3)}{2\sqrt{2}}}
\e^{i\frac{\phi(\eta_4)}{2\sqrt{2}}}
\prod_{i=1}^N \e^{i\sqrt{2}\phi(z_i)} \right \ra . \qquad
\eeqa
Next we use the well-known formula
\[
\left \la \prod_{a=1}^N \e^{i \lambda_a \phi(z_a)} \right \ra =
\prod_{a<b} \left(z_a -z_b \right)^{\lambda_a\lambda_b}
\delta_{\lambda_1+\cdots \lambda_N,0}
\]
to show that
\[
\left \la \e^{i\frac{1}{\sqrt{2}}\phi(\eta_2)}
\e^{i\frac{1}{2\sqrt{2}}\phi(\eta_3)}
\e^{i\frac{1}{2\sqrt{2}}\phi(\eta_4)}
\prod_{i=1}^N \e^{i\sqrt{2}\phi(z_i)} \right \ra =
\eta_{23}^{1/4} \eta_{24}^{1/4} \eta_{34}^{1/8}
\left \la
\begin{array}{c}\mathrm{independent \ of} \cr \eta_{ab}\end{array}
\right \ra ,
\]
where the second expectation value on the right-hand side is independent
of $\eta_{ab}$
and therefore gives no contribution to monodromies.
Similarly, noting in passing that the branch-cut structure of the wave
functions
with 2 or 3 quasiholes is determined by the corresponding 2-pt and 3-pt
 quasihole functions, for even $N$ we find
\[
 \left \la \s (\eta_3)\s(\eta_4)\prod_{i=1}^N \psi(z_i) \right \ra   =
 \eta_{34}^{-1/8}
 \left \la
\begin{array}{c}\mathrm{single-valued  } \cr  \mathrm{in} \
\eta_{ab}\end{array}
\right \ra  \qquad \mathrm{and}
\]
\[
 \left \la \psi(\eta_2)\s (\eta_3)\s(\eta_4)\prod_{i=1}^N \psi(z_i) \right \ra   =
 \frac{\eta_{34}^{3/8}}{\eta_{23}^{1/2} \eta_{24}^{1/2}}
 \left \la
\begin{array}{c}\mathrm{single}-\mathrm{valued } \cr  \mathrm{in}\ \eta_{ab}\end{array}
\right \ra .
\]
Thus we finally obtain, for even number $N$ of electrons,
\beqa
\Psi_0 &=& \left(\eta_{23}\eta_{24}\right)^{\frac{1}{4}}
\left \la
\begin{array}{c}\mathrm{single-valued} \cr \mathrm{in} \ \eta_{ab}\end{array}
\right \ra \nn   \label{wf-1-2}
\Psi_1 &=& \frac{\sqrt{\eta_{12}\eta_{34}}}{\left(\eta_{23}\eta_{24}\right)^{\frac{1}{4}} }
\left \la
\begin{array}{c}\mathrm{single-valued} \cr \mathrm{in} \  \eta_{ab}\end{array}
\right \ra  .
\eeqa
We point out that the factors containing $\eta_{ab}$, hence the monodromies
of the wave functions (\ref{wf-1-2}), are independent
of whether the absolute value of $\tx$ is bigger or smaller than 1 because
considering $\eta_{12}\to 0$ actually means $|\tx| < 1$.

Now the read-out procedure, Eq.~(\ref{read-out}), which after the fusion
$\eta_1\to \eta_2$ is reduced to $\eta_{23}\to \e^{2\pi i}\eta_{23}$,
 simply gives for the channel passing through P and Q in Fig.~\ref{fig:qubit}
\[
\Psi_0 \to \e^{i\frac{\pi}{2}} \Psi_0, \quad   \mathrm{while}\quad
\Psi_1 \to \e^{-i\frac{\pi}{2}} \Psi_1
\]
so that the diagonal conductance is in agreement with Eqs.~(\ref{sigma_0})
and  (\ref{sigma_1})
\beq \label{sigma1-2}
\sigma_{xx} \propto |t_{MN}+ i\,  t_{PQ}|^2 \ \mathrm{for}\ \Psi_0, \quad
\mathrm{while} \quad
\sigma_{xx} \propto  |t_{MN}- i\,  t_{PQ}|^2 \ \mathrm{for}\ \Psi_1 .
\eeq
\section{The NOT gate of Das Sarma et al.}
\label{sec:NOT}
The original idea of Ref.~\cite{sarma-freedman-nayak,freedman-nayak-walker}
behind  the construction of the logical NOT gate in the setup shown on
Fig.~\ref{fig:qubit}, is that when
the quasiholes with coordinates $\eta_1$ and $\eta_2$, localized on
antidots 1 and 2, form the qubit \cite{nayak-private}
another quasihole  with coordinate $\eta_3$ could
tunnel between the front gates A and B (through an additional
antidot located between A and B in order to guarantee a single
quasihole tunneling) in such a way to flip  the state of the qubit.
The point is that this third quasihole actually encircles the
quasihole localized
on antidot 1 but not the second quasihole localized on antidot 2.
More generally, while the read-out of the qubit is performed by taking
$\eta_3$ around $\eta_1$ and $\eta_2$, the NOT gate could be executed
by taking $\eta_3$ around $\eta_1$ or $\eta_2$ but not both of them.

To understand this in more detail let us consider the simplest situation
when the quasiholes are ordered $|\eta_1|>|\eta_2|>|\eta_3|>|\eta_4|$
and the quasihole at $\eta_3$ traverses a closed loop around the quasihole
at $\eta_2$, i.e.,
\beq \label{NOT}
\mathrm{\bf NOT}:\qquad \eta_{23} \rightarrow \lim_{t\to 1_-}
\e^{ 2\pi i \, t}\eta_{23},
\quad \mathrm{where} \quad  t\in [0, 1)
\eeq
with all other quasiholes coordinates remaining unchanged.
Provided that  the three quasiholes at $\eta_1$, $\eta_2$ and
$\eta_3$ are kept well-separated,
after the tunneling of the third quasihole the 4-qh wave functions
(\ref{01}) are adiabatically transformed into new wave functions
which we shall find now.
First the transformation (\ref{NOT}) transforms the crossratio (\ref{x})
according to $x \to\e^{2\pi i}x$. Next we note that the Pfaffian wave
functions $\Psi_{(ab)(cd)}$ are single-valued, so that they remain
unchanged  under  the NOT transformation (\ref{NOT}).
Finally, the square root of $x$ changes sign under the NOT transformation
(\ref{NOT}), i.e.,
\[
\sqrt{x} \to \sqrt{\e^{2\pi i} x}= \e^{i\pi } \sqrt{x} .
\]
The transformation of $\sqrt{1\pm \sqrt{x}}$ is more subtle and depends on
the absolute value of the crossratio $x$. When $|x|<1$ we have
$\sqrt{1\pm \sqrt{x}} \to \sqrt{1\mp \sqrt{x}}$ under the transformation
(\ref{NOT}), while if $|x|>1$ $\sqrt{1\pm \sqrt{x}} \to \pm\sqrt{1\mp \sqrt{x}}$,
 because the transformation contour now encircles both branching points.
Thus
we find that the two 4-qh wave functions (\ref{01}) transform under the
NOT transformation (\ref{NOT}) as follows
\beqa \label{NOT-1}
U^{\mathrm{Das \ Sarma}}_{\bf NOT}
\left(\matrix{\Psi_{4\mathrm{qh}}^{(0)} \cr  \Psi_{4\mathrm{qh}}^{(1)} } \right)  & = &
\left(\matrix{0 & 1 \cr 1 & 0}\right)
\left(\matrix{\Psi_{4\mathrm{qh}}^{(0)} \cr  \Psi_{4\mathrm{qh}}^{(1)} } \right)
 \quad \quad \mathrm{for} \quad |x|<1, \quad \mathrm{while} \\    \label{NOT-2}
U^{\mathrm{Das \ Sarma}}_{\bf NOT}
\left(\matrix{\Psi_{4\mathrm{qh}}^{(0)} \cr  \Psi_{4\mathrm{qh}}^{(1)}} \right)  & = &
\left(\matrix{\quad 0 & 1 \cr -1 & 0}\right)
\left(\matrix{\Psi_{4\mathrm{qh}}^{(0)} \cr  \Psi_{4\mathrm{qh}}^{(1)} } \right)
\quad \mathrm{for} \quad |x|>1    .
\eeqa
Thus we conclude that the transformation (\ref{NOT}) indeed maps
$\Psi_{4\mathrm{qh}}^{(0)}\leftrightarrow\Psi_{4\mathrm{qh}}^{(1)}$.
\section{Exchange matrices and monodromy group representation 
of the Ising 4-point functions}
\label{sec:R4}
The four-point correlation functions of the chiral spin field (\ref{sigma})
in the Ising model can be shown to be \cite{fst}
\beqa \label{F}
F(\eta_1,\eta_2,\eta_3,\eta_4) &\equiv &
\la \sigma(\eta_1)\sigma(\eta_2)\sigma(\eta_3)\sigma(\eta_4) \ra=
F_+ + F_-, \quad \mathrm{where} \nn
F_{\pm}(\eta_1,\eta_2,\eta_3,\eta_4) &=& \frac{1}{\sqrt{2}}
\left(\frac{\eta_{13}\eta_{24}}{\eta_{12}\eta_{14}\eta_{23}\eta_{34}}\right)^{\frac{1}{8}}
\sqrt{1\pm \sqrt{\frac{\eta_{14}\eta_{23}}{\eta_{13}\eta_{24}}}}
\eeqa
are the chiral conformal blocks, which could be expressed in terms of the 
fields $\sigma_\pm$ with definite fermion parity as 
$F_\pm=\la \sigma_+ (\eta_1) \sigma_\pm (\eta_2)
\sigma_+(\eta_3)\sigma_\pm(\eta_4) \ra$.

\begin{figure}[htb]
\centering
\includegraphics*[bb=130 460 430 700,width=8cm]{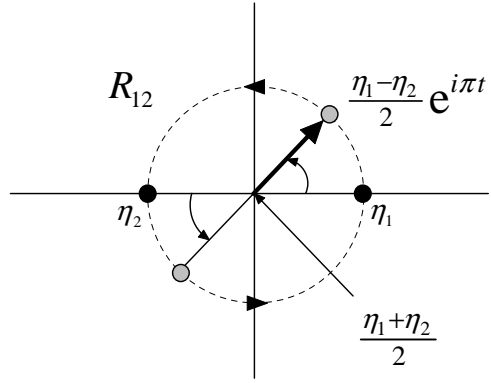}
 \caption{Counter-clockwise exchange $R_{12}$ of coordinates
 	$\eta_1$ and $\eta_2$ performed  by taking the limit $t=0\to 1_-$ in
	Eq.~(\ref{contour_12}).}
\label{fig:R_12}
\end{figure}

The counter-clockwise exchange of the quasiparticles with coordinates $\eta_1$
and $\eta_2$ can be preformed by analytic continuation along the contour 
defined by
\beq \label{contour_12}
\eta'_1= \frac{\eta_1+\eta_2}{2} + \e^{i\pi t } \frac{\eta_1-\eta_2}{2},  \quad
\eta'_2= \frac{\eta_1+\eta_2}{2} - \e^{i\pi t } \frac{\eta_1-\eta_2}{2} , \quad
0 \leq t \leq 1,
\eeq
as shown in Fig.~\ref{fig:R_12}.
Executing this transformation and taking the limit $t\to 1_{-}$ we obtain for
the 4-pt functions (note that $\eta'_{12}= \e^{i\pi} \eta_{12}$)
\[
R_{12}F_\kappa =  \left(
\frac{\e^{-i\pi}\eta_{13}\eta_{24}}{\eta_{12}\eta_{14}\eta_{23}\eta_{34} }
\right)^{\frac{1}{8}}
\left( \frac{\eta_{14}\eta_{23}}{\eta_{13}\eta_{24}}\right)^{\frac{1}{4}}
\sqrt{1+\kappa \sqrt{\frac{\eta_{13}\eta_{24}}{\eta_{14}\eta_{23}}}} =
\e^{-i\frac{\pi}{8}}  \sqrt{\kappa}F_\kappa (\eta_a)  .
\]
Therefore (fixing the signs of the square roots by the requirement that the
$R$-matrices must satisfy the Yang--Baxter equations (\ref{artin}))
we get  a diagonal matrix
\[
R_{12} \left(\matrix{ F_+ \cr F_- }\right) = \e^{-i\frac{\pi}{8}}
\left( \matrix{1 & 0 \cr 0 & i}\right)
\left(\matrix{ F_+ \cr F_-} \right)  .
\]
Precisely in the same way we can compute the exchange matrix  $R_{34}$
which takes precisely the same form in this basis.
In order to compute the exchange matrix $R_{23}$ we apply the coordinate
transformation (\ref{contour_12}), after renaming
$(\eta_1,\eta_2) \to (\eta_2,\eta_3)$, and use the identity
\[
\sqrt{1+\sqrt{x}} +i\lambda \sqrt{1-\sqrt{x}} =
\sqrt{2\lambda i }\sqrt{\sqrt{1-x} -i\lambda \sqrt{x}}
\]
for $\lambda=\pm 1$. We obtain
\[
R_{23} F_\pm =\frac{ \e^{-i\frac{\pi}{8}}}{\sqrt{2}}
\left(
\frac{\eta_{13}\eta_{24}}{\eta_{12}\eta_{14}\eta_{23}\eta_{34} }
\right)^{\frac{1}{8}}
\sqrt{\sqrt{1-x} \pm i \sqrt{x}} =  \frac{ \e^{-i\frac{\pi}{8}}}{\sqrt{2}}
 \e^{\pm i\frac{\pi}{4}} \left(F_+ \mp i F_- \right) .
\]
Thus, we can summarize our results for the elementary exchange matrices
 $R_{a, a+1}$ in the basis $\{F_+,F_-\}$
\beq \label{R-4pt}
R_{12}=R_{34}=\e^{-i\frac{\pi}{8}} \left( \matrix{1 & 0 \cr 0 & i}\right),
\quad
R_{23}=\frac{\e^{i\frac{\pi}{8}} }{\sqrt{2}}
\left( \matrix{1 & -i \cr -i & 1}\right).
 \eeq
All other exchanges can be obtained from the elementary ones, e.g.,
\[
R_{13} =  R_{12}^{-1} R_{23} R_{12}=
\frac{\e^{i\frac{\pi}{8}} }{\sqrt{2}} \left( \matrix{\ \ \ 1 & 1 \cr -1 &  1}\right) ,
\quad \mathrm{etc.}
\]
\subsection{Exchange matrices for the 4-quasiholes wave functions
	$\Psi_{4\qh}^{(0,1)}$ }
The 4-quasiholes wave functions (\ref{01}) are built up from the 4-point
functions of the Ising model and the functions $\Psi_{(ab)(cd)}$ which are
single-valued in the positions of the quasiholes. Now a natural question
arises: is the braid-group representation for the 4-point functions
(\ref{f}) extended to a braid-group representation over the 4-quasihole
functions? The answer is  yes.
The point is that when a quasihole traverses a closed loop, the
4-quasihole functions should acquire an additional phase proportional to the
number of electrons inside the loop (or, to the area of the loop).
This might lead to a projective representation of the braid group as looks to be
the case in the   the $p$-wave composite-fermion superconductor approach, where
the quasihole is identified with a half-quantum vortex \cite{ivanov}.
However, as can be seen in our approach directly from the 4-quasihole functions
(\ref{01}), each electron inside the loop contributes $2\pi$ to this phase
because the quasiholes are by definition local with the
electrons\footnote{what could really change  the phase is not  the entire
electron but the neutral Majorana fermion, which is non-local with the
quasihole}. Thus the phase is insensitive to the number of electrons inside
the loop, it only counts the number of quasiholes. Nevertheless it turns out
that braid-group (or mapping class group) \cite{birman} representations in
terms of CFT correlation functions are generically projective.
The point is that the coordinates of the many-body wave functions,
which in the CFT approach \cite{5-2} are chiral correlation functions
defined on the unit circle, could be naturally extended by analytic 
continuation
to the vicinity of the unit disk. Then by CFT transformation these functions
could be extended to the entire \textit{compactified} complex plane,
which is isomorphic to the two-dimensional sphere. Now, besides the Artin
relations  \cite{birman} 
\beqa \label{artin}
	B_i B_j &=&  B_j B_i, \qquad \qquad \mathrm{for}  \quad |i-j|\geq 2 \nn
  B_i B_{i+1} B_i &=&  B_{i+1} B_i B_{i+1}, \quad \mathrm{where}
	\quad B_i=R_{i,i+1} \in \B_n,
\eeqa
for the generators $B_i$ ($i=1,\ldots, n-1$) of $\B_n$,
the representation of the braid group on the sphere should satisfy
 one more relation \cite{birman}
\[
	B_1 B_2 \cdots B_{n-2} B_{n-1}^2 B_{n-2} \cdots  B_2 B_1 = \I.
\]
As can be seen by direct computation the above relation  is satisfied
by the elementary exchange matrices  of the chiral CFT correlators only
up to phase so that the braid-group representation is projective.

In the rest of this subsection we will show that from the exchange matrices
for the functions $F_\pm$ we could obtain the corresponding matrices in the
basis $\{ \Psi_{4\qh}^{(0)}, \  \Psi_{4\qh}^{(1)} \}$ in the form 
\cite{TQC-PRL}
\beq \label{R4}
R_{12}^{(4)}=R_{34}^{(4)}=\left( \matrix{1 & 0 \cr 0 & i}\right),
\quad
R_{23}^{(4)}=
\frac{\e^{i\frac{\pi}{4}} }{\sqrt{2}}
\left( \matrix{\ \    1 & -i \cr -i & \ \ 1}\right),
\eeq
where the superscript ``$(4)$'' is to remind us that these matrices are
computed in the basis of the 4-quasiholes states (\ref{01}).
Consider, for instance, the transformation (\ref{contour_12}) acting on the
4-quasiholes wave functions (\ref{01}). It changes $x\to 1/x$ and
$\Psi_{(13)(24)}\leftrightarrow \Psi_{(14)(23)}$  so that
\[
R_{12}^{(4)}\Psi_{4\qh}^{(0,1)} =
\frac{(\eta_{23}\eta_{14})^{\frac{1}{4}}}{\sqrt{1\pm\sqrt{\frac{1}{x}}}}
\left(\Psi_{(14)(23)} \pm \frac{1}{\sqrt{x}} \Psi_{(13)(24)}\right)  .
\]
Expressing $\eta_{23}\eta_{14}=\eta_{13}\eta_{24} \, x$ and taking out
$\sqrt{\pm \sqrt{1/x}}$ from  the denominator we obtain the exchange matrix
$R_{12}^{(4)}$ in the basis (\ref{01}) as in Eq.~(\ref{R4}).
Notice that there is no more $\e^{-i\pi/8}$ in this matrix.
At this point the Yang--Baxter equations (\ref{artin}) imply that if there is a braid-group
representation over the 4-quasihole wave functions (\ref{01}) they must be
obtained from Eq.~(\ref{R-4pt}) by multiplying all exchange matrices with
$\e^{i\pi/8}$. Because $R_{12}^{(4)}$   is diagonal it could be
directly obtained by first fusing $\eta_1\to \eta_2$ and then interpreting
the exchange as $\eta_{12}\to \e^{i\pi}\eta_{12}$, which gives the same result.
Indeed, it follows from Eq.~(\ref{wf-1-2}) that when
$\eta'_1 = \eta_2 -\e^{i\pi} \eta_{12}$, $\eta'_2=\eta_2$, which is equivalent to
$R_{12}^{(4)}$ the two functions $\Psi_0$ and  $\Psi_1$ in Eq.~(\ref{wf-1-2})
are multiplied by 1 and $i$ respectively (note that the expression in the
$\la \cdots \ra$  in Eq.~(\ref{wf-1-2})
is actually independent of $\eta_1$).

Next, instead of computing $R_{23}^{(4)}$ directly it is more convenient to
compute  $R_{13}^{(4)}$, following Ref.~\cite{nayak-wilczek}, and then use
the identity
\beq \label{R_23-id}
R_{23}^{(4)}=R_{12}^{(4)}R_{13}^{(4)}\left(R_{12}^{(4)}\right)^{-1}.
\eeq
To this end we first apply the coordinate transformation
\[
	\eta'_1=\eta_3, \quad  \eta'_2=\eta_2,  \quad \eta'_3=\eta_1 , \quad \eta'_4=\eta_4  ,
	\quad   \mathrm{such \ that}  \quad
	\eta'_{13}=\e^{i\pi}\eta_{13},
\]
$x \to  \tx=1-x$
and $\Psi_{(13)(24)}\to \Psi_{(13)(24)}$, while $\Psi_{(14)(23)}\to \Psi_{(12)(34)}$.
Then, using the Nayak--Wilczek identity \cite{nayak-wilczek}
(note the misprints in Eq.~(3.8) there)
\[
	(1-x) \Psi_{(12)(34)} =  \Psi_{(13)(24)}  - x \, \Psi_{(14)(23)} ,
	\quad \mathrm{as \ well \ as}
\]
\beq \label{sqrt-id}
\sqrt{1+\sqrt{1-\tx}} \pm \sqrt{1-\sqrt{1-\tx}} = \pm \sqrt{2} \sqrt{1\pm \sqrt{\tx}},
\eeq
it is not difficult to derive $R_{13}^{(4)}$, hence obtain $R_{23}^{(4)}$
by Eq.~(\ref{R_23-id}), i.e.,
\[
R_{13}^{(4)} \left( \matrix{\Psi_{4\qh}^{(0)} \cr \Psi_{4\qh}^{(1)}}\right) =
\frac{\e^{i\frac{\pi}{4}}}{\sqrt{2}}
\left( \matrix{\ \ \ 1 & 1 \cr -1 & 1}\right)
\left( \matrix{\Psi_{4\qh}^{(0)} \cr \Psi_{4\qh}^{(1)}}\right)    .
\]
The sign ambiguity in front of the square root in
the right-hand side of Eq.~(\ref{sqrt-id})  comes from taking a square root
and is not directly linked to the sign under the square root on the right.
In order to fix this sign we use Eq.~(\ref{R_23-id}) and the fact that
the NOT gate $U^{\mathrm{Das \ Sarma}}_{\bf NOT}$ for $|x|<1$, that we obtained
in Eq.~(\ref{NOT-1}) directly in terms of the 4-quasiholes wave functions
monodromies in  Sect.~\ref{sec:NOT},
is actually the square of  the exchange matrix  $R_{23}^{(4)}$, i.e.,
\[
\left(R_{23}^{(4)}\right)^2 \Psi_{4\qh}^{(0,1)}= \Psi_{4\qh}^{(1,0)} \quad
\Rightarrow \quad
\left(R_{23}^{(4)}\right)^2=\left( \matrix{0 & 1 \cr 1 & 0}\right).
\]
\begin{rem}
The relevance of the Pfaffian qubit for quantum computation can be
emphasized once again.  Because the monodromy
matrix $\left(R_{23}^{(4)}\right)^2$ coincides with  the {\rm NOT} gate in
the 4-quasiholes states basis,  the matrix $R_{23}^{(4)}$ should be identified
with $\sqrt{\mathrm{NOT}}$, which cannot be implemented in classical
information theory \cite{sqrt-NOT}.  As we shall see later this operation is
crucial for the construction of the Hadamard gate, which is one of the most
important single-qubit quantum gate.
\end{rem}
\begin{rem}
The derivation of the elementary exchange matrices (\ref{R4})) generating the 
entire two-dimensional representation of the braid group $\B_4$, over the 4-quasiholes 
Pfaffian wave functions, follows the lines of Ref.~\cite{nayak-wilczek} where the 
first row of the exchange matrix $R_{13}^{(4)}$ has been explicitly   computed. 
These matrices can be obtained from the general representation of the braid group $\B_4$
for the Pfaffian state, as derived in Ref.~\cite{slingerland-bais},
using the quantum group structure of the $\Z_k$ parafermion Hall states,
 see Eqs. (138) and (140) there. Nevertheless, the direct and self-contained derivation of 
the exchange matrices, given above, from the analytic properties of the  Pfaffian wave functions has 
certain advantages and provides an independent check of the results.
\end{rem}

The monodromy transformations, corresponding to the elementary exchanges, i.e.,
the squares of $R_{a, a+1}^{(4)}$ representing a complete counter-clockwise cycle
of the particle with label $a+1$ around that with label $a$,
\[
\left(R_{12}^{(4)}\right)^2=\left(R_{34}^{(4)}\right)^2=
\left( \matrix{1 & \ \ \ 0 \cr 0 & -1}\right),
	\quad
 \left(R_{23}^{(4)}\right)^2= \left( \matrix{0 & 1 \cr 1 & 0}\right)  ,
\]
form a subgroup of the braid group called the monodromy group.
The monodromy group representation is thus generated by the Pauli
matrices $\sigma_1$ and $\sigma_3$, or, alternatively, by the two
elements\footnote{note that the second Pauli matrix here appears
in this context naturally  multiplied by $i$ and this is in accord 
with the standard quantum computation conventions \cite{nielsen-chuang}}
$S=i\sigma_2$, $R= \sigma_1$, such that $S^4=R^2=\I$ and $R^{-1}SR=S^{-1}$.
Therefore the image of the monodromy group is isomorphic to the finite
non-Abelian group $\mathcal{D}_4$ \cite{coxeter-moser,hamermesh},  
known as the symmetry 
group of the square, which has $8$ elements typically given in the 
two-dimensional representation as
\[
\mathcal{D}_4\equiv \{ \pm \I_2, \ \pm \sigma_1 ,  \ \pm i\sigma_2 , \ 
\pm \sigma_3 \},
\]
where $\sigma_i$ are the three Pauli matrices.
The monodromy group $\mathcal{M}_n$ is in general a normal subgroup of the 
braid group $\B_n$ and the factor-group
\[
\B_n/\mathcal{M}_n \simeq \mathcal{S}_n
\]
is isomorphic to the permutation group $\mathcal{S}_n$. Therefore, the fact that the
monodromy group representation $\mathcal{M}_4=\mathcal{D}_4$ for the 4-pt functions is finite
implies that the braid group representation in this case is a finite group
whose order could be shown to be
$|\mathrm{Image}\left(\B_4\right)| = 96$. The order of the representation of
the braid group $\B_6$ is
$|\mathrm{Image}\left(\B_6\right)|=46080 = 2^6 6!$,
while that of the monodromy subgroup is
$|\mathrm{Image}\left(\mathcal{M}_6\right)|=32$. Similarly, the order of the 
representation of the braid group $\B_8$ is
$|\mathrm{Image}\left(\B_8\right)| = 5160960=2^7 8!$, and 
that of its monodromy subgroup is
$|\mathrm{Image}\left(\mathcal{M}_8\right)|=128$.
These numbers have been obtained by direct enumeration of
the distinct matrices, produced by multiplying the elementary braid matrices,
in the corresponding braid-group representation using the Dimino's
algorithm \cite{dimino}.
In general the image of the representation of the braid group $\B_{2n}$,
for $n\geq 3$, over the $2n$-point functions of the Ising model is
\cite{read-JMP}
\[
|\mathrm{Image}\left(\B_{2n}\right)| = \left\{
\begin{array}{lll}2^{2n-1} (2n) ! & \quad \mathrm{for} & n=\mathrm{even} \\
2^{2n} (2n) ! & \quad \mathrm{for} & n=\mathrm{odd} \end{array} \right. .
\]
While this may look very nice as a mathematical fact it is fairly
disappointing from the perspective of topological quantum computation. The
point is that our intention in TQC is to implement quantum gates by simply
exchanging quasiparticles positions and the finite braid-group representation
 implies that we could generate only finite number of gates with the
Pfaffian qubit. Therefore it cannot be used for universal TQC where we
would like to efficiently implement any unitary matrix (with a given
precision).
However, the set of quantum gates that could be realized by braiding of
Pfaffian quasiparticles is known to be a Clifford group, which plays a
central role in quantum error correction codes, so their topologically
protected construction is fairly important.
Moreover, in the next Section we will try to circumvent this restriction
by finding a unitary transformation, which does not belong to the braid group,
that could be used to construct a universal set of gates,
and is topologically protected.

\subsection{The read-out of the single-qubit state in terms of the exchange
matrices}
\begin{figure}[htb]
\centering
\includegraphics*[bb=20 350 560 480,width=\textwidth]{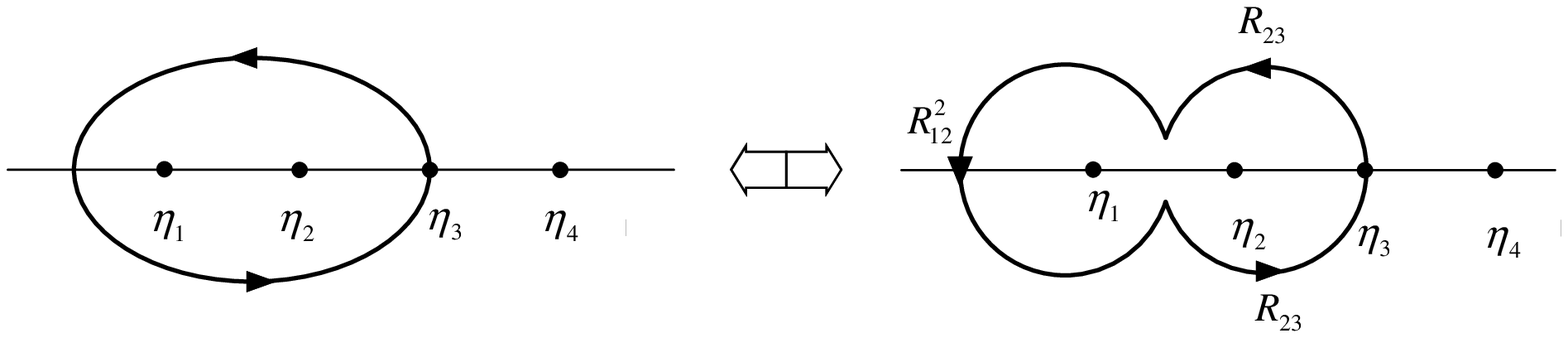}
 \caption{The read-out for $|\tx|<1 $ in terms of the elementary  exchange
 matrices.}
\label{fig:read-out-R}
\end{figure}

\begin{figure}[htb]
\centering
\includegraphics*[bb=25 380 560 480,width=\textwidth]{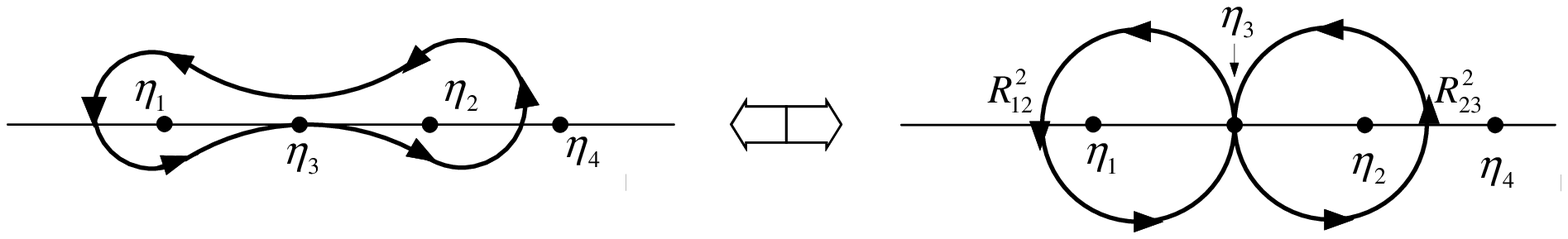}
 \caption{The read-out for $|\tx|>1$ in terms of the elementary
exchange matrices.}
\label{fig:read-out-R-2}
\end{figure}

Instead of interpreting the read-out as ``taking $\eta_3$ around $\eta_4$'',
as we did in Sects.~\ref{sec:read-out-34-1} and \ref{sec:read-out-34-2}, we
could use the exchange matrices (\ref{R4}) to
 directly  ``take $\eta_3$ around $\eta_1$ and $\eta_2$''.
Again the measurement result depends on whether the absolute value of the
crossratio (\ref{tx}) is smaller or bigger than 1.
In the case $|\tx|<1$, which is the usual situation when, e.g.,
$|\eta_1| >|\eta_2| >|\eta_3| >|\eta_4| $,
as shown on Fig.~\ref{fig:read-out-R}, the operator
corresponding to the transportation of $\eta_3$ around $\eta_1$ and
$\eta_2$ is
\[
U^{(|\tx|<1)}_{\mathrm{read-out}} = R_{23} R_{12}^2 R_{23} =
\frac{\e^{i\frac{\pi}{2}} }{2}
\left( \matrix{\ \ 1 & -i \cr -i & \ \ 1}\right)
\left( \matrix{1 & \ \ 0 \cr 0 & -1}\right)
\left( \matrix{\ \ 1 & -i \cr -i & \ \ 1}\right)    =
\left( \matrix{i & \ \ 0 \cr 0 & -i}\right),
\]
and the result is the same as Eq.~(\ref{U_<}). If on the other hand
$|\tx|>1$,  which can be considered as, e.g.,
$|\eta_2| < |\eta_3|$, because using the the CFT symmetry we can express
$\tx=\eta_{3}/\eta_2$. In this case the read-out operation corresponds to
 Fig.~\ref{fig:read-out-R-2},
is
\[
U^{(|\tx|>1)}_{\mathrm{read-out}} =    R_{12}^2 R_{23}^2=
\left( \matrix{1 & \ \ 0 \cr 0 & -1}\right)
\left( \matrix{0 & 1 \cr 1 & 0}\right)
 = \left( \matrix{\ \ 0 & 1 \cr -1 & 0}\right)
\]
and this is (up to phase) identical with  Eq.~(\ref{U_>}).
\section{Single-qubit gates constructed from elementary exchange
matrices}
\label{sec:single-qubit}
In this Section we shall implement by quasihole braiding the following
single-qubit gates: the Hadamard gate $H$, Pauli gates $X$, $Y$, $Z$, and
the phase gate $S$ \cite{nielsen-chuang}). In Sect.~\ref{sec:CNOT}
we shall also construct entirely in terms of quasihole braidings
 the Controlled-NOT gate.
While not sufficient for universal QC, these gates are known to form
a Clifford group  \cite{bravyi-kitaev-Clifford-QC} and could in principle
be used for universal quantum computation provided that we can create
the so-called  \textit{magic states} \cite{bravyi-kitaev-Clifford-QC}.
The only single-qubit gate
that we cannot construct directly by braiding  is the  $\pi/8$ gate $T$
\cite{nielsen-chuang}). However, instead of the $\pi/8$ gate one
we could use  the three-qubit Toffoli gate \cite{nielsen-chuang}).

The Hadamard gate $H$ is of central importance for any QC scheme.
It is worth stressing that the Hadamard gate is the only gate
which must be added to a universal classical computer (based on
the Toffoli gate) in order to make it a universal quantum computer
\cite{H-T-universal}.
In  the TQC with Pfaffian qubits it can be used to create
special superpositions, called the Bell states (or EPR states)
\cite{nielsen-chuang}, that can be constructed in no other way.
In addition $H$ is one of the building blocks of the quantum
Fourier transform  \cite{nielsen-chuang}).
The Hadamard gate in the Pfaffian TQC  scheme can be expressed
in terms of three elementary braidings of
 the 4-quasiholes states (\ref{01}), namely \footnote{the results in this 
Section  have  been originally announced in Ref.~\cite{TQC-PRL}; here we 
give a detailed derivation},
\beq \label{H}
H\simeq R_{12}^2 R_{13} = R_{12} R_{23} R_{12}  =
\frac{\e^{i\frac{\pi}{4}} }{\sqrt{2}}
\left( \matrix{1 & \ \ \  1 \cr 1 & -1}\right).
\eeq
\begin{figure}[htb]
\centering
\includegraphics*[bb=0 330 595 520,width=12cm]{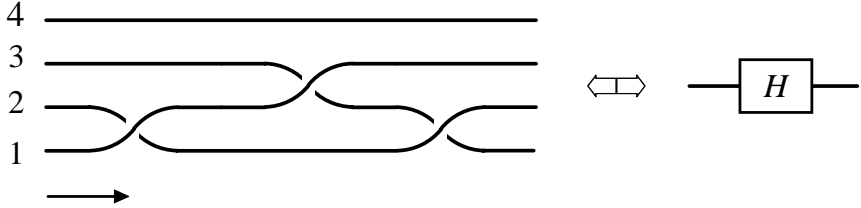}
 \caption{Braiding diagram for the Hadamard gate (\ref{H}) and its symbol
 	(on the right)  in standard	quantum-computation notation}
\label{fig:H}
\end{figure}
It would be convenient to represent these braidings in braid diagrams,
such as Fig.~(\ref{fig:H}). The first and the third braids
in Fig.~(\ref{fig:H}) are
in clockwise direction and correspond to the inverse exchanges $R_{12}^{-1}$,
while the second exchange is in counter-clockwise direction and
corresponds to $R_{23}$.

The Pauli $X$ gate, known also as the NOT gate,  which was first implemented
for the Pfaffian qubit in Ref.~\cite{sarma-freedman-nayak}, could be executed by two
elementary exchanges as
\[
X \equiv  R_{23}^2=  \left( \matrix{0 & 1 \cr 1 & 0}\right)
\]
\begin{figure}[htb]
\centering
\includegraphics*[bb=0 340 595 520,width=9cm]{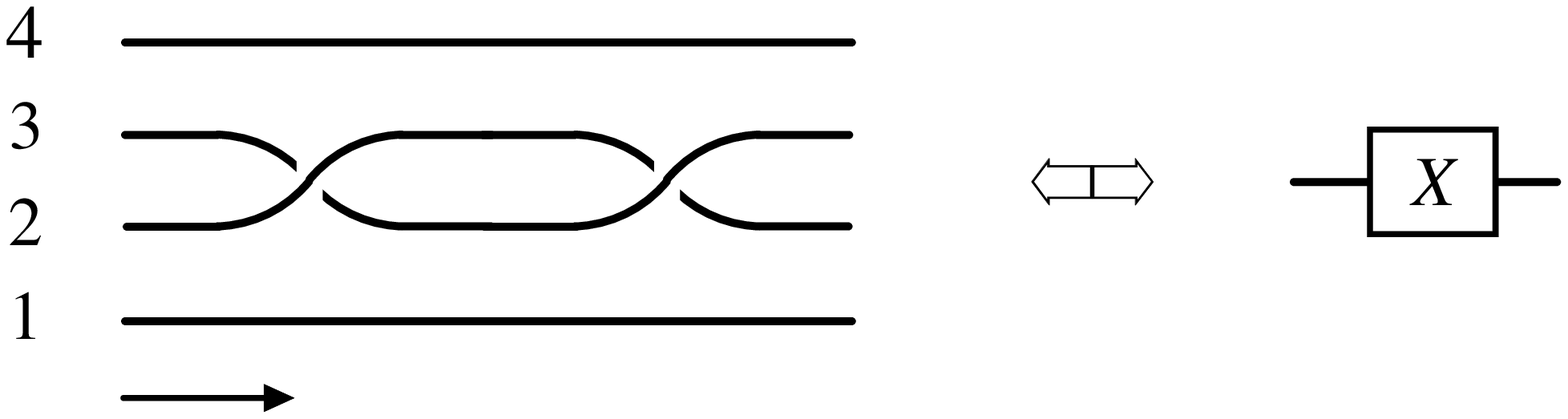}
 \caption{Braiding diagram for the Pauli $X$ gate and its
 	quantum-computation symbol}
\label{fig:X}
\end{figure}
and the corresponding braid diagram is shown on Fig.~\ref{fig:X}.
Similarly, the $Y$ gate, which is usually defined in quantum computation
literature without the imaginary unit, is realized by  4 elementary
exchanges as follows
\[
Y\equiv R_{12}^{-1}R_{23}^2 R_{12}=
\left( \matrix{\ \ \ 0 & 1 \cr -1 & 0}\right)
\]
and the corresponding braid diagram is shown on Fig.~\ref{fig:Y}.
\begin{figure}[htb]
\centering
\includegraphics*[bb=0 330 595 530,width=12cm]{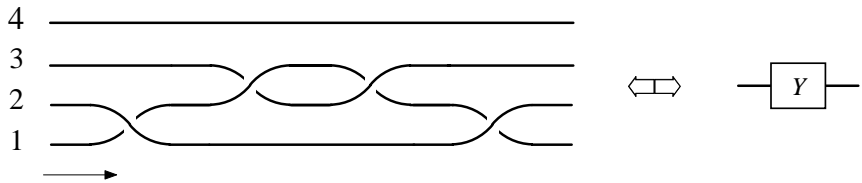}
 \caption{Braiding diagram for the Pauli $Y$ gate}
\label{fig:Y}
\end{figure}
The Pauli $Z$ gate \cite{nielsen-chuang} can be realized in two
different ways by two elementary exchanges as
\[
Z\equiv R_{12}^2= R_{34}^2 =
\left( \matrix{1 & \ \ \ 0 \cr 0 & -1}\right)
\]
and the braid diagram for the  first of them is shown on Fig.~\ref{fig:Z}.
\begin{figure}[htb]
\centering
\includegraphics*[bb=0 330 595 530,width=12cm]{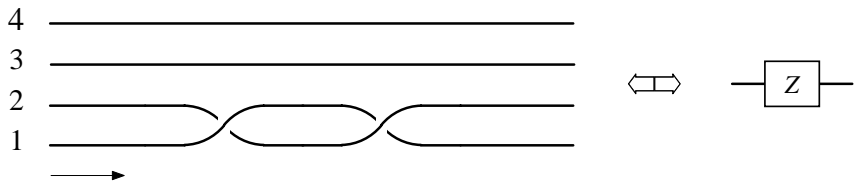}
 \caption{Braiding diagram for the Pauli $Z$ gate}
\label{fig:Z}
\end{figure}
Finally, the  phase gate $S$, which can also be realized in two different ways
by a single elementary exchange,
\[
S\equiv R_{12}= R_{34} = \left( \matrix{1 & 0 \cr 0 & i}\right)
\]
and the first of them, $S=R_{12}$, is shown graphically on Fig.~\ref{fig:S}.
\begin{figure}[htb]
\centering
\includegraphics*[bb=0 330 595 530,width=12cm]{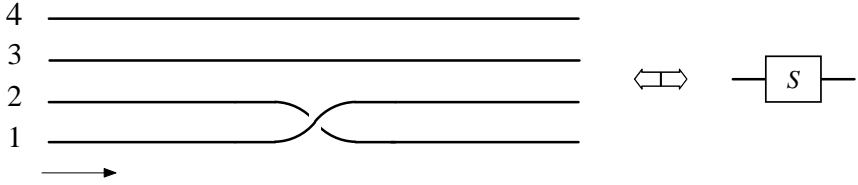}
 \caption{Braiding diagram for the phase  gate  $S$}
\label{fig:S}
\end{figure}
Notice that $S^2=Z$ as it should be.

The only single-qubit gate which cannot be implemented directly in terms of quasiholes
braiding is the $\pi/8$-gate $T=\mathrm{diag}\left(1,\e^{i\pi/4}\right)$
\cite{nielsen-chuang} because $\det T =\e^{i\pi/4}$, while
$\det \left(R_{a,a+1}^{(4)}\right)=i$ for all $a$.
Instead of $T$ we shall propose to construct the Toffoli gate.
\section{Two-qubits construction and two-qubit gates}
\label{sec:2qubits}
In order to realize two qubits, which belong to $\C^2$, we need at
least 4-dimensional space. Recalling that the dimension of the excited Pfaffian
states with $2n$ quasiholes at fixed positions   \cite{nayak-wilczek} is
dim $\H_{2n} =2^{n-1}$ we consider the 6-quasihole states.
\begin{figure}[htb]
\centering
\includegraphics*[bb=80 570 510 700,width=12cm]{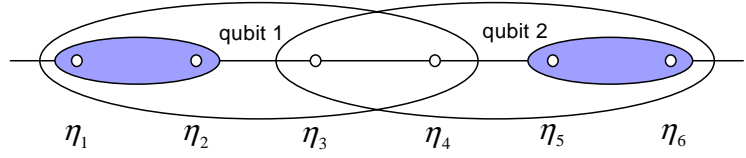}
 \caption{Two qubits, realized by quasiholes with coordinates
$(\eta_1,\eta_2,\eta_3,\eta_4)$ and 	$(\eta_3,\eta_4,\eta_5,\eta_6)$,
spanning the needed 4 dimensional space $\C^2$. The state of qubit 1 is
determined by the quasiholes with positions $(\eta_1,\eta_2)$, while the
state of qubit 2 by the quasiholes with positions $(\eta_5,\eta_6)$ and that
is why these two groups were shaded. The state of the quasiholes with
coordinates $(\eta_3,\eta_4)$ depends on both quasihole pairs
$(\eta_1,\eta_2)$ and $(\eta_5,\eta_6)$.}
\label{fig:2qubits}
\end{figure}
Before we explain how to construct the two-qubit states let us recall that
the single qubit states can be written as
\[
|0\ra\equiv \la \s_+ \s_+\s_+\s_+\ra, \quad
|1\ra\equiv \la \s_+ \s_-\s_+\s_-\ra,
\]
where we take the convention that the first two $\s$
 fields determine the state of the qubit while the last two guarantee
the preservation of the fermion parity, so that basically
$|0\ra \simeq \s_+ \s_+$, while $|1\ra \simeq \s_+ \s_-$, which is in agreement
with our definition of the qubit because of the fusion rules
$\s_+ \s_+ \sim \I$ and  $\s_+ \s_- \sim \psi$.

The two-qubit basis is defined here by the convention that the first
two quasiholes form the first qubit while the last two quasiholes form
the second qubit
\beqa \label{2qb-basis}
&&|00\ra \equiv \la \s_+ \s_+\s_+\s_+\s_+\s_+\ra, \quad
\! |01\ra \equiv \la \s_+ \s_+\s_+\s_-\s_+\s_-\ra \nn
&&|10\ra \equiv \la \s_+ \s_-\s_+\s_-\s_+\s_+\ra, \quad
|11\ra \equiv \la \s_+ \s_-\s_+\s_+\s_+\s_-\ra  .
\eeqa
This is convenient since if we fuse the first two quasiholes this would
project to the second qubit, while if we fuse the last two quasiholes
this would project to the first qubit, i.e.,
\beq \label{project}
|\alpha\beta\ra \mathop{\to}_{\eta_1\to\eta_2} |\beta\ra , \quad
|\alpha\beta\ra \mathop{\to}_{\eta_5\to\eta_6} |\alpha\ra .
\eeq
Then the third and the fourth quasiholes are fixed by the conservation
of the fermion parity, i.e., if $e_i$ is the parity of $\s_{e_i}$,
consider the correlation function with $\gamma_F$ plugged in the middle.
This gives
\[
\la \s_{e_1} \s_{e_2}\s_{e_3} \gamma_F\s_{e_4}\s_{e_5}\s_{e_6}\ra=
e_1e_2e_3\la \s_{e_1} \s_{e_2}\s_{e_3}\s_{e_4}\s_{e_5}\s_{e_6}\ra
\]
if we move $\gamma_F$ to the left or
$e_4e_5e_6\la \s_{e_1} \s_{e_2}\s_{e_3}\s_{e_4}\s_{e_5}\s_{e_6}\ra$
if we move it to the right. Therefore we obtain the fermion parity rule
\[
 e_1 e_2 e_3=e_4e_5e_6 \quad \Rightarrow \quad   e_3e_4=e_1e_2e_5e_6,
\]
where the last equality follows from $e_i^2=1$.
Thus we have only 4 independent states in the space of 6-quasiholes
with fixed positions, which correspond to Eq.~(\ref{2qb-basis}).
\subsection{Exchange matrices for 6-quasiholes}
The braid group representation over the 6-pt functions is generated by the
elementary exchanges $R^{(6)}_{12}$,  $R^{(6)}_{23}$,   $R^{(6)}_{34}$,
$R^{(6)}_{45}$ and $R^{(6)}_{56}$. We shall construct them explicitly by
using the operator-product expansions of the Ising model and the projections
to the single-qubit states along the lines of Ref.~\cite{TQC-PRL}.

The construction (\ref{2qb-basis}) of the two-qubit states allows to derive
the 6-quasiholes exchange matrices in terms of those for the 4 quasiholes.
In order to obtain the exchange  $R^{(6)}_{12}$ we may first fuse
$\eta_5 \to \eta_6$,
because the state of the first qubit is independent of $\eta_5$ and $\eta_6$,
and use the projections (\ref{project}). In more detail, we have
\beqa \label{fuse_56}
&&|00\ra \mathop{\to}_{\eta_5 \to \eta_6}  \la \s_+ \s_+\s_+\s_+\ra, \quad
|01\ra \mathop{\to}_{\eta_5 \to \eta_6}   \la \s_+ \s_+\s_+\s_- \psi\ra \nn
&&|10\ra \mathop{\to}_{\eta_5 \to \eta_6}  \la \s_+ \s_-\s_+\s_-\ra, \quad
|11\ra \mathop{\to}_{\eta_5 \to \eta_6}  \la \s_+ \s_-\s_+\s_+\psi \ra  .
\eeqa
Now the exchange $\eta_1 \leftrightarrow \eta_2$ is represented  by the action
of $R_{12}^{(4)}$, which we take  from Eq.~(\ref{R4}), i.e.,
\beqa
&&|00\ra \mathop{\to}_{\eta_1 \leftrightarrow \eta_2}   \ \ \la \s_+ \s_+\s_+\s_+\ra, \quad
|01\ra \mathop{\to}_{\eta_1 \leftrightarrow \eta_2}  \ \  \la \s_+ \s_+\s_+\s_- \psi\ra \nn
&&|10\ra \mathop{\to}_{\eta_1 \leftrightarrow \eta_2} i \ \la \s_+ \s_-\s_+\s_-\ra, \quad
|11\ra \mathop{\to}_{\eta_1 \leftrightarrow \eta_2} i \ \la \s_+ \s_-\s_+\s_+\psi \ra   \nonumber
\eeqa
and restoring back the second qubit we obtain in the basis (\ref{2qb-basis})
\beq \label{R6_12}
	R_{12}^{(6)} =  \left(
\matrix{1 & 0 & 0 & 0 \cr  0 & 1 & 0 & 0 \cr 0 & 0 & i & 0 \cr 0 & 0 & 0 & i} \right)
=R^{(4)}_{12} \otimes \I_2 .
\eeq
Now let us compute the exchange  matrix $R^{(6)}_{23}$. Again we can
fuse $\eta_5 \to \eta_6$  and use Eq.~(\ref{fuse_56}).
The exchange $\eta_2 \leftrightarrow \eta_3$ is represented  by the action
of $R_{23}^{(4)}$,  from Eq.~(\ref{R4}), i.e.,
\beqa
&&|00\ra \mathop{\to}_{\eta_2 \leftrightarrow \eta_3}
\ \  \frac{\e^{i\frac{\pi}{4}} }{\sqrt{2}}\la \s_+ \s_+\s_+\s_+\ra -
i\  \frac{\e^{i\frac{\pi}{4}} }{\sqrt{2}}\la \s_+ \s_-\s_+\s_-\ra, \nn
&&|01\ra \mathop{\to}_{\eta_2 \leftrightarrow \eta_3}  \ \
\frac{\e^{i\frac{\pi}{4}} }{\sqrt{2}}\la \s_+ \s_+\s_+\s_- \psi\ra  -
i\  \frac{\e^{i\frac{\pi}{4}} }{\sqrt{2}}\la \s_+ \s_-\s_+\s_+ \psi\ra, \nn
&&|10\ra \mathop{\to}_{\eta_2 \leftrightarrow \eta_3}  \
 \frac{\e^{i\frac{\pi}{4}} }{\sqrt{2}}\la \s_+ \s_-\s_+\s_-\ra
 -i\ \frac{\e^{i\frac{\pi}{4}} }{\sqrt{2}}\la \s_+ \s_+\s_+\s_+\ra, \nn
&&|11\ra \mathop{\to}_{\eta_2 \leftrightarrow \eta_3}
\frac{\e^{i\frac{\pi}{4}} }{\sqrt{2}}\la \s_+ \s_-\s_+\s_+\psi \ra - i \
\frac{\e^{i\frac{\pi}{4}} }{\sqrt{2}}\la \s_+ \s_+\s_+\s_-\psi \ra  \nonumber
\eeqa
and restoring back again the second qubit according to Eq.~(\ref{fuse_56})
we obtain in the basis (\ref{2qb-basis})
\beq \label{R6_23}
	R_{23}^{(6)} = \frac{\e^{i\frac{\pi}{4}} }{\sqrt{2}} \left(
\matrix{1 & 0 & -i & 0 \cr  0 & 1 & 0 & -i \cr -i & 0 & 1 & 0 \cr 0 & -i & 0 & 1} \right)
=R^{(4)}_{23} \otimes \I_2 .
\eeq
Precisely in the same way, by fusing first $\eta_1 \to \eta_2$, we can obtain the
6-quasiholes  exchange matrices
\beq \label{R6_45}
R^{(6)}_{45}  =\frac{\e^{i\frac{\pi}{4}} }{\sqrt{2}}\left(
\matrix{1 & -i  & 0 & 0 \cr  -i & 1 & 0 & 0 \cr 0 & 0 & 1 & -i \cr 0 & 0 & -i & 1} \right)=
\I_2 \otimes R^{(4)}_{23} , \quad \mathrm{and}
\eeq
\beq \label{R6_56}
R^{(6)}_{56}  = \left(
\matrix{1 & 0 & 0 & 0 \cr  0 & i & 0 & 0 \cr 0 & 0 & 1 & 0 \cr 0 & 0 & 0 & i} \right)
= \I_2 \otimes R^{(4)}_{34} .
\eeq
The exchange  matrix $R^{(6)}_{34}$ cannot be obtained simply in this way
because the quasiholes at $\eta_3$ and $\eta_4$ depend on the states of both
qubits so that fusing either $(\eta_1,\eta_2)$ or $(\eta_5,\eta_6)$ would change the state
of the quasihole pair at $(\eta_3,\eta_4)$.
As we shall see in Sect.~\ref{sec:CZ} this entanglement of the two qubits
allows us to construct immediately the Controlled-Z gate.
Notice, however, that  $R^{(6)}_{34}$ must be diagonal and therefore could be
directly determined by simply using the OPE for the fusion $\eta_3 \to \eta_4$ 
alone.
One way to see this is that if $R^{(6)}_{34}$ were non-diagonal the exchange of quasiholes at
$\eta_3$  and $\eta_4$ would create a coherent superposition of the states
$\I |\mathrm{NS}\ra$ and $\psi |\mathrm{NS}\ra$ in the Neveu--Schwartz  sector
(states with even number of $\s$ fields acting on the vacuum belong to the NS sector)
which would be a violation of the superselection rule  defined by the chiral fermion parity
$\gamma_F$. In contrast, if there are odd number of $\s$'s to the right of
$\eta_a$ and $\eta_{a+1}$ acting on the vacuum then the fusion  $\eta_a \to\eta_{a+1}$
generates $\I |\mathrm{R}\ra$ and $\psi |\mathrm{R}\ra$, which are in the Ramond sector
where the chiral fermion parity is spontaneously broken \cite{5-2}, so that the above
states could indeed form coherent superpositions. This explains why  $R^{(6)}_{23}$ and
$R^{(6)}_{45}$ could be non-diagonal, while  $R^{(6)}_{12}$,  $R^{(6)}_{34}$ and
 $R^{(6)}_{56}$  have to be diagonal.
Thus, using the (neutral part of the) OPE (\ref{qh-OPE}), we have
\beqa \label{fuse_34}
&&|00\ra \mathop{\to}_{\eta_3 \to \eta_4}  \eta_{34}^{-1/8}\la \s_+ \s_+\s_+\s_+\ra, \quad
|01\ra \mathop{\to}_{\eta_3 \to \eta_4}   \eta_{34}^{3/8}\la \s_+ \s_+ \psi \s_+\s_- \ra \nn
&&|10\ra \mathop{\to}_{\eta_3 \to \eta_4}  \eta_{34}^{3/8}\la \s_+ \s_- \psi \s_+\s_+\ra, \quad
|11\ra \mathop{\to}_{\eta_3 \to \eta_4}  \eta_{34}^{-1/8}\la \s_+ \s_-\s_+\s_- \ra  .
\eeqa
Therefore, the exchange  $\eta_3 \leftrightarrow \eta_4$,  which simply transforms
$\eta_{34} \to \e^{i\pi}\eta_{34}$, leads to
\beqa
&&|00\ra \mathop{\to}_{\eta_3 \leftrightarrow \eta_4}
\ \  \e^{-i\frac{\pi}{8}} \eta_{34}^{-1/8}\la \s_+ \s_+\s_+\s_+\ra , \nn
&&|01\ra \mathop{\to}_{\eta_3 \leftrightarrow \eta_4}  \ \
 i  \ \e^{-i\frac{\pi}{8}}  \eta_{34}^{3/8}\la \s_+ \s_+ \psi \s_+\s_- \ra \nn
&&|10\ra \mathop{\to}_{\eta_3 \leftrightarrow \eta_4}  \ \
 i \ \e^{-i\frac{\pi}{8}}\eta_{34}^{3/8}\la \s_+ \s_- \psi \s_+\s_+\ra, \nn
&&|11\ra \mathop{\to}_{\eta_3 \leftrightarrow \eta_4}   \ \
\e^{-i\frac{\pi}{8}} \eta_{34}^{-1/8}\la \s_+ \s_-\s_+\s_- \ra . \nonumber
\eeqa
Taking into account that there is another $\eta_{34}^{1/8}$, coming from the $u(1)$
component of the quasihole operator,
we find  the exchange  matrix $R^{(6)}_{34}$  in the basis (\ref{2qb-basis})
to be
\beq \label{R6_34}
R^{(6)}_{34}  = \left(
\matrix{1 & 0 & 0 & 0 \cr  0 & i & 0 & 0 \cr 0 & 0 & i & 0 \cr 0 & 0 & 0 & 1} \right) .
\eeq
Unlike the other 6 quasiholes exchange matrices, $R^{(6)}_{34}$ is
not a factorized tensor product  of  the 4 quasiholes exchange matrices.
Instead, there is an additional  built-in structure in this representation
of the braid group, which will eventually allow us to construct the
Controlled-$Z$ gate.

It is easy to check that the 6-quasiholes exchange matrices (\ref{R6_12}),
(\ref{R6_23}), (\ref{R6_45}), (\ref{R6_56}) and (\ref{R6_34}) indeed satisfy
the Artin relations (\ref{artin}) for the braid group $\B_6$ \cite{birman},
including the Yang--Baxter equations.
As mentioned before, using the Dimino's algorithm \cite{dimino} we can 
explicitly obtain 
the entire group generated by the matrices (\ref{R6_12}), (\ref{R6_23}), 
(\ref{R6_34}), (\ref{R6_45}) and (\ref{R6_56}), giving the orders of the 
representation of the braid group $\B_6$ and its monodromy subgroup
\[
|\mathrm{Image}(\B_6)|=46080, \quad |\mathrm{Image}(\mathcal{M}_6)|=32.
\]
\subsection{The Controlled-$Z$ gate in terms of 6-quasiholes braidings}
\label{sec:CZ}
Using the explicit expressions for the 6-pt $R$-matrices, Eqs.~(\ref{R6_12}),
(\ref{R6_34}) and (\ref{R6_56}), it is straight forward to construct the
most important two-qubit gates---the CNOT or CZ gates in terms of the
braid matrices, e.g.,
\beq \label{true-CZ}
	\mathrm{CZ} = \  R_{12}^{(6)} \left(R_{34}^{(6)}\right)^{-1}  R_{56}^{(6)}   =
\left(
\matrix{1 & 0 & 0 & \ \ \ 0 \cr 0 & 1 & 0  & \ \ \ 0 \cr 0 & 0 & 1 & \ \ \ 0
	\cr 0 & 0 & 0 &  -1}
\right)                .
\eeq
Some insight into the CZ construction may be
gained from the identity \cite{nielsen-chuang}
\[
 \mathrm{CZ} =\e^{i\frac{\pi}{4}}  \e^{i\frac{\pi}{4} Z_1 Z_2} \e^{-i\frac{\pi}{4}Z_1}
	\e^{-i\frac{\pi}{4}Z_2},
\]
where $Z_1$ and $Z_2$ are the $Z$ gates acting on qubit 1 and 2 respectively.
Because these matrices are diagonal and square to $\I$ it is not difficult to prove that
their exponents are actually equal (up to overall phases) to
the matrices  $R_{12}^{(6)}$, $R_{56}^{(6)}$ respectively,   while the exponent of
$Z_1Z_2$ is proportional to the inverse of  $R^{(6)}_{34}$.

\begin{figure}[htb]
\centering
\includegraphics*[bb=0 330 595 520,width=14cm]{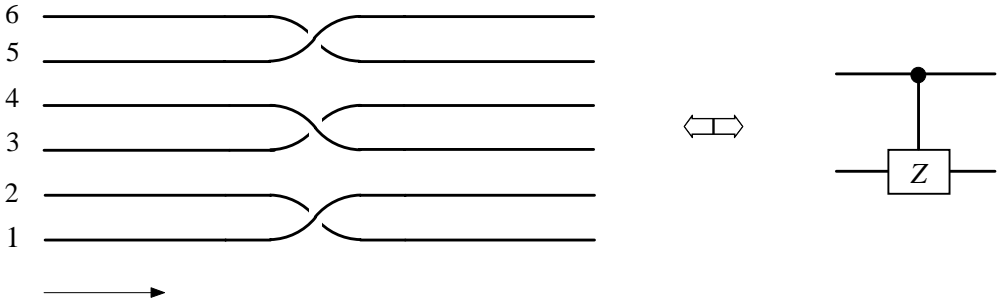}
\caption{The braid diagram for the  Controlled-Z gate realized by
3 commuting elementary 6-quasiparticle braidings defined in
Eq.~(\ref{true-CZ}). The symbol on the right is the standard
quantum-computation  notation for CZ.}
\label{fig:CZ-braid}
\end{figure}

The  braid diagram for the 6-quasihole exchanges corresponding to Eq.~(\ref{true-CZ})
is shown on Fig.~\ref{fig:CZ-braid}.
In plotting Fig.~\ref{fig:CZ-braid} we have used that the three $R$-matrices
entering Eq.~(\ref{true-CZ}) are diagonal and therefore commute, which
also follows from the Artin relations (\ref{artin}), so that the order of the exchanges
is not important.
Note the remarkable simplicity of this realization of the CZ
gate---just three elementary exchanges. This is one of the main advantages
of the two-qubit construction in terms of 6 quasiholes presented in
Sect.~\ref{sec:2qubits}.
\subsection{Single-qubit gates in the two-qubit basis}
\label{sec:single-qubit}
Before we continue, it is important to show that we can efficiently express
the single-qubit gates into the two-qubit basis (\ref{2qb-basis}).
The point is that the exchange matrices for 4 quasiholes, which represent the
single-qubit operations, belong to the braid group
$\B_4$ while those for the two-qubit gates are expressed in terms of braid matrices from
the braid group $\B_6$ and the former have different structure from the latter.
Physically, the embedding of the one-qubit gates into the two-qubit system is
non-trivial because the entanglement creates non-local effects between the two qubits.
Nevertheless, the single-qubit construction in terms of 4 quasiholes exchanges is
certainly instructive for the representation of these gates in the two-qubit basis.
For our purposes it would be convenient to construct these gates explicitly.

The Hadamard gate acting on the first qubit can be expressed as
\[
H_1 \simeq H  \otimes \I_2 = \left(R_{12}^{(6)}\right)^{-1} \left(R_{23}^{(6)}\right)^{-1}
 \left(R_{12}^{(6)}\right)^{-1}= \frac{\e^{-i\frac{\pi}{4}}}{\sqrt{2}} \left(
\matrix{1 & 0 & \ \ \ 1 & \ \ \ 0 \cr 0 & 1 & \ \ \ 0  & \ \ \ 1
	\cr 1 & 0 & - 1 &  \ \ \ 0 	\cr 0 & 1 & \ \ \ 0 &  -1}
\right) ,
\]
while that acting on the second qubit should be identified with
\beq \label{H_2}
H_2 \simeq \I_2 \otimes H = R_{56}^{(6)} R_{45}^{(6)}  R_{56}^{(6)}=
\frac{\e^{i\frac{\pi}{4}}}{\sqrt{2}} \left(
\matrix{1 & \ \ \ 1 & 0 & \ \ \ 0 \cr 1 & -1 & 0  & \ \ \ 0 \cr 0 & \ \ \ 0 & 1 & \ \ \ 1
	\cr 0 & \ \ \ 0 & 1 &  -1}
\right) .
\eeq
Both Hadamard gates have similar  structures to their single-qubit
counterparts, yet, they are slightly different.
This is surprising because adding an additional qubit is equivalent to 
introducing two more strands in the braid diagram and we expect that the two
straight lines representing a trivial braiding in the new qubit should 
correspond to the the unit operator in a tensor product with the single-qubit 
gate acting on the old qubit. The point is that, however, the representation 
of the braid group $\B_6$ realized by the 6-quasihole Pfaffian wave functions 
naturally appear in a different basis, which is not a factorized tensor 
product of the representations of $\B_4$ over the 4-quasiholes Pfaffian wave 
functions. This is some kind of topological entanglement which seems to be 
common for all TQC schemes based on non-Abelian anyons realized in FQH systems.

The phase gates acting on the first and second qubits are respectively
\[
	S_1=S\otimes \I_2= R_{12}^{(6)} \quad \mathrm{and} \quad
	S_2=\I_2 \otimes S= R_{56}^{(6)} .
\]
The embeddings of the other single-qubits into the two-qubit basis follow
from these of $H$ and $S$ because $Z=S^2$ and $HZH=X$.
\subsection{The Controlled-NOT gate}
\label{sec:CNOT}
Now that we know how to construct the Controlled-Z gate, and how to embed the
single-qubits gates,  entirely in terms of 6-quasiholes braidings, the
CNOT gate is readily computed with the help of the target-qubit Hadamard
gate (\ref{H_2}), i.e.,
\beq \label{CNOT}
\mathrm{CNOT}=
H_2 \  \mathrm{CZ} \ H_2
 =  R_{56} R_{45} R_{56}^{-1} R_{34}^{-1} R_{12}R_{45} R_{56}    \simeq
 \left(
\matrix{1 & 0 & 0 &  0 \cr 0 & 1 & 0  &  0 \cr 0 & 0 & 0 & 1
	\cr 0 & 0 & 1 &  0}
\right) 	.
\eeq
We used here that the target-qubit Hadamard gate could be executed by
3 exchanges, according to Eq.~(\ref{H_2}),
as well as the property $\left(R_{56}\right)^4=\I_4$ and some of the Artin
relations (\ref{artin}) for the 6-quasiholes exchange matrices.
An equivalent realization of CNOT, which gives precisely the same result
as in Eq.~(\ref{CNOT}), could be given in terms of other 7 elementary
exchanges
\beq \label{CNOT-2}
\mathrm{CNOT}=R_{34}^{-1} R_{45} R_{34}R_{12}R_{56} R_{45}  R_{34}^{-1} .
\eeq
\begin{figure}[htb]
\centering
\includegraphics*[bb=10 340 580 490,width=\textwidth]{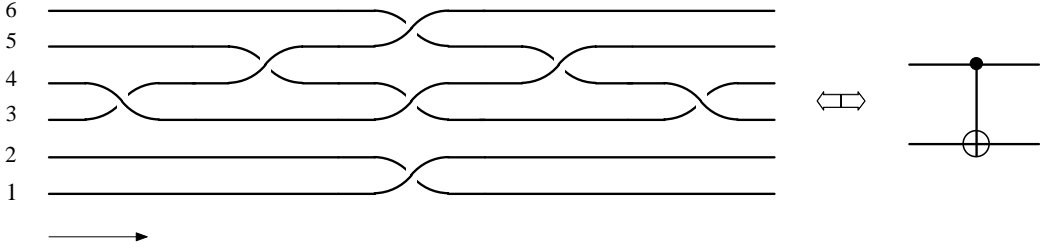}
\caption{The braid diagram for the  Controlled-NOT gate executed by 7
elementary 6-quasiparticle braidings corresponding to Eq.~(\ref{CNOT-2}).
The symbol on the right is the standard quantum-computation  notation
for CNOT.}
\label{fig:CNOT-braid}
\end{figure}
This is the first known construction of the Controlled-Z and
Controlled-NOT gates
entirely in terms of the braid matrices for 6 quasiholes in the Pfaffian
TQC scheme, which certainly guarantees  the exactness and topological
protections of  these gates.

Note that this construction of the CNOT gate is equivalent to the braid 
realization of the Bell matrix of Refs.~\cite{kauffman-YB,kauffman-braid}.
The algebraic structure behind this Bell matrix, when it is used as a 
universal $R$ matrix  in the $R(T\otimes T) = (T\otimes T) R$ relations, 
giving rise to an exotic new bialgebra called $S03$,
 has been clarified in Ref.~\cite{arnaudon}. 
\subsection{The Bravyi--Kitaev Controlled-$Z$ gate precursor}
\label{sec:CZ-pre}
\begin{figure}[htb]
\centering
\includegraphics*[bb=40 360 550 500,width=\textwidth]{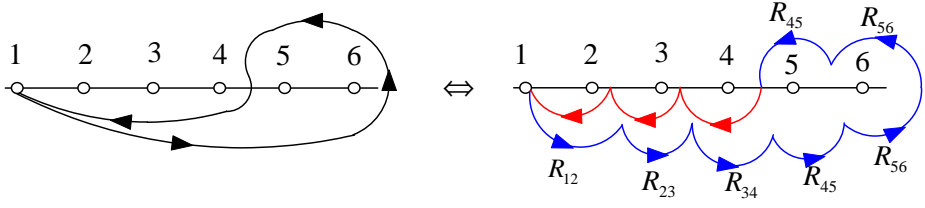}
 \caption{The Controlled-Z gate precursor in terms of the monodromies
 	$\widetilde{\mathrm{CZ}}=R_{15}^2 R_{16}^2$.}
\label{fig:C-Z}
\end{figure}

\begin{figure}[htb]
\centering
\includegraphics*[bb=50 490 540 630,width=\textwidth]{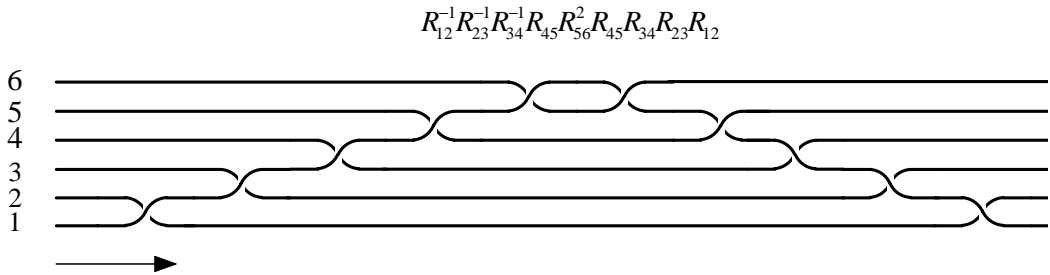}
 \caption{The braid diagram for the monodromy-based Controlled-Z gate precursor.}
\label{fig:CZ-braid0}
\end{figure}

For the sake of completeness and comparison we shall also describe the
existing idea \cite{freedman-nayak-walker} to realize CZ by taking one
quasihole, around two other, which suffers from the drawback that the resulting
transformation has one extra minus sign and thus has to be supplemented by external
operations in order to produce the CZ gate, see below.
Consider, e.g., the quasihole with position $\eta_1$, from qubit 1
transported adiabatically around the two
quasiholes, with positions $\eta_5$ and $\eta_6$, of qubit 2
(or, equivalently, taking the two quasiholes comprising
qubit 2 around one quasihole of qubit 1) as shown on Fig.~\ref{fig:C-Z}.
This is obviously equivalent to first taking $\eta_1$  around $\eta_6$ and
then around $\eta_5$
so that this gate would be just the product of the two corresponding monodromies
\beq \label{CZ}
	\widetilde{\mathrm{CZ}}= R_{15}^2  R_{16}^2 \simeq
	R_{12}^{-1}R_{23}^{-1} R_{34}^{-1} R_{45}R_{56}^2 R_{45}R_{34}R_{23}R_{12} =
	\left(
\matrix{1 & 0 & 0 & 0 \cr 0 & \!\!\!\! -1 & 0  & 0 \cr 0 & 0 & 1 & 0
	\cr 0 & 0 & 0 & \!\!\!\! -1} \right)  .
\eeq
The second equality in Eq.~(\ref{CZ}) is just an equivalent representation
which can be readoff from Fig.~\ref{fig:C-Z}.
We give for convenience also the explicit expression for $R_{15}^2$
\[
	R_{15}^2=   R_{12}^{-1} R_{23}^{-1} R_{34}^{-1}  R_{45}R_{34}R_{23} R_{12}  =
\e^{i\frac{\pi}{4}}
\left(
\matrix{0 & 0 & 0 & 1 \cr 0 & 0 &  -1  & 0 \cr 0 & 1 & 0 & 0 \cr
	-1 & 0 & 0 &  0}
\right)  ,
\]
which together with Eq.~(\ref{R2-16}) below can be used to compute
$\widetilde{\mathrm{CZ}}$ directly.
The braid diagram for this realization
of the CZ gate precursor is shown on Fig.~\ref{fig:CZ-braid0}.
Now it is obvious that the gate we constructed in Eq.~(\ref{CZ}) differs
from the CZ gate by having one more minus sign. The reason is that if the
 second qubit is in the state $|1\ra$ the transport of the quasihole at $\eta_1$
 will always produce a minus sign whatever the state of the first qubit
 (note that the two states which are multiplied by $-1$ by our gate (\ref{CZ})
 are exactly $|01\ra$ and $|11\ra$).
 The idea of Bravyi--Kitaev \cite{freedman-nayak-walker} is to split the first
 qubit into two  $1/4$-charge states only if this qubit is in the state $|1\ra$
 and then move the two quasiholes
 (at positions $\eta_5$ and $\eta_6$ in our case) forming the second qubit around
 the first qubit in order to execute this gate.
 This would remove the minus sign from the second row of
 $\widetilde{\mathrm{CZ}}$ in Eq.~(\ref{CZ}) and, if successfully implemented,
   should give us a topologically   protected  Controlled-Z gate .
\subsection{Realization of the Bravyi--Kitaev two-qubit	gate $g_3$}
\label{sec:g_3}
One particular universal set of quantum gates, relevant for TQC with
Pfaffian qubits, which has been  proposed by Bravyi and Kitaev, is
\cite{freedman-nayak-walker}
\[
g_1=\left(\matrix{1 & 0 \cr 0 & \e^{i\frac{\pi}{4}}} \right), \ \
g_2=\left(
\matrix{1 & 0 & 0 & 0 \cr 0 & 1 & 0  & 0 \cr 0 & 0 & 1 & 0 \cr
	0 & 0 & 0 &  \!\!\!\! -1} \right),
\ \
g_3=\frac{1}{\sqrt{2}}\left(
\matrix{1 & 0 & 0 & -i \cr 0 & 1 & -i  & 0 \cr 0 & -i & 1 & 0 \cr -i & 0 & 0 &  1} \right).
\]
The two-qubit gate $g_2$ is identical with our Controlled-Z gate
(\ref{true-CZ})
implemented in a topologically protected manner by 6-quasiholes braidings.
The single-qubit gate $g_1$, known also as the $\pi/8$ gate $T$,
has been realized in Ref.~\cite{freedman-nayak-walker}
as an unprotected gate, by bringing together the two quasiholes for a short
period of time and then pulling them back,  in which the exponential
topological protection is lost.

\begin{figure}[htb]
\centering
\includegraphics*[bb=40 360 550 500,width=\textwidth]{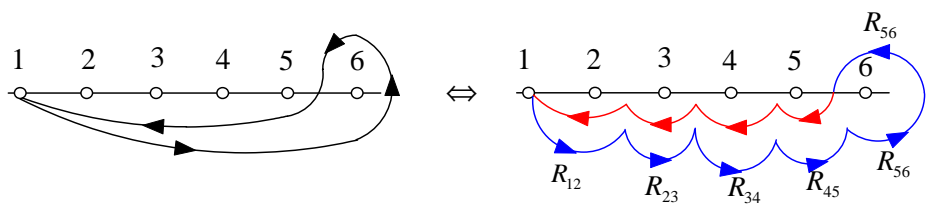}
\caption{The monodromy $R_{16}^2$ executed by taking $\eta_1$ around
$\eta_6$ in a counter-clockwise direction.}
\label{fig:R2_16}
\end{figure}

\begin{figure}[htb]
\centering
\includegraphics*[bb=40 490 520 620,width=\textwidth]{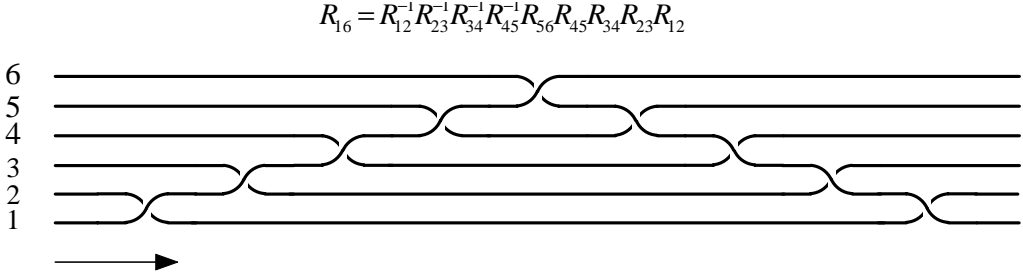}
\caption{Implementing the two-qubit gate $g_3$ by counter-clockwise exchange
of the quasiholes with positions $\eta_1$ and $\eta_6$.}
\label{fig:R_16}
\end{figure}

In trying to construct the gate $g_3$ it would be instructive to compute first
the monodromy $R_{16}^2$ as taking $\eta_1$ around $\eta_6$. As is obvious from
Fig.~\ref{fig:R2_16}, this monodromy can be expressed in terms of the
elementary 6-quasiholes exchanges\footnote{note that the clockwise exchanges
correspond to the inverse exchange matrices}  (omitting the superscript
(6) in the  $R$-matrices) as
\beq \label{R2-16}
 R_{16}^2 = R_{12}^{-1} R_{23}^{-1} R_{34}^{-1} R_{45}^{-1}
	 R_{56}^2 R_{45}R_{34}R_{23}R_{12} =
\left(
\matrix{0 & 0 & 0 & 1 \cr 0 & 0 & 1  & 0 \cr 0 & 1 & 0 & 0 \cr 1 & 0 & 0 &  0}
\right) .
\eeq
 Then the exchange $R_{16}$ is just the ``square-root''  of
Eq.~(\ref{R2-16}), i.e.,
\beq \label{R_16}
R_{16} = R_{12}^{-1} R_{23}^{-1} R_{34}^{-1} R_{45}^{-1}
R_{56} R_{45}R_{34}R_{23}R_{12} .
\eeq
The direct computation, using the explicit formulas,
Eqs.~(\ref{R6_12}), (\ref{R6_23}),  (\ref{R6_34}), (\ref{R6_45}) and
(\ref{R6_56}) for the elementary $R$ matrices,  shows that the two-qubit
gate $g_3$  identical with $R_{16}$, i.e.,
\[
R_{16} \simeq \frac{1}{\sqrt{2}}
\left(
\matrix{1 & 0 & 0 & -i \cr 0 & 1 & -i  & 0 \cr 0 & -i & 1 & 0 \cr -i & 0 & 0 &  1} \right)
  \equiv g_3    .
\]
The braid diagram for the realization of $R_{16}$ is shown in
Fig.~\ref{fig:R_16}.

\begin{rem}
It is worth stressing that the TQC scheme of Das Sarma et al.
\cite{sarma-freedman-nayak} is essentially
based on the monodromy transformations of the multi-quasihole Pfaffian wave 
functions. The only exception is the  Bravyi--Kitaev gate $g_3$, 
which was proposed to be constructed schematically 
in Ref.~\cite{freedman-nayak-walker} 
by braiding two among 4 quasiholes (not by braiding 6 quasiholes as it 
should be). Note, however, that as a two-qubit gate acting on Pfaffian qubits, 
$g_3$ must be constructed in terms of transformations of the 6-quasiholes 
states, which has not been done in Ref.~\cite{freedman-nayak-walker}.
In particular, it is not clear from the 4-quasihole braiding construction of 
Ref.~\cite{freedman-nayak-walker}, which two quasiholes among the 6 ones 
must be exchanged in order to obtain the gate $g_3$, and whether  
this is at all possible. 
The first explicit results proving the usefulness of braidings for 
topological quantum computation with Pfaffian quasiholes  have been 
obtained  in Ref.~\cite{TQC-PRL} and Eq.~(\ref{R_16}) is the  
braiding implementation of $g_3$.
\end{rem}

Just for reference, and to demonstrate the importance\footnote{note that 
$R_{25}$ and $R_{16}$ are not equivalent as quantum operations} of the 
choice of 
quasiholes to be exchanged, we note that in a similar way  the 
exchange of  $\eta_2$ with $\eta_{5}$ gives rise to the following matrix
\[
R_{25} =   R_{23}^{-1} R_{34}^{-1}  R_{45}R_{34}R_{23}   =
\frac{\e^{i\frac{\pi}{4}}}{\sqrt{2}}
\left(
\matrix{1 & 0 & 0 & -i \cr 0 & 1 & i  & 0 \cr 0 & i & 1 & 0 \cr -i & 0 & 0 &  1}
\right).
\]
\subsection{The non-demolition measurement gate }
\label{sec:non-demolition}
One of the quantum gates in the universal TQC schemes reviewed
in Ref.~\cite{freedman-nayak-walker} is executed by a non-demolition
measurement of the total topological charge of two qubits.
According to our construction of the two-qubit states this measurement
is equivalent to the transformation
\[
|00\ra \to |00\ra, \quad
|01\ra \to -|01\ra, \quad
|10\ra \to -|10\ra, \quad
|11\ra \to |01\ra .
\]
Therefore we can identify the non-demolition measurement gate with
the monodromy matrix corresponding to the adiabatic transport of
$\eta_3$ around $\eta_4$
\[
\left(R_{34}^{(6)}\right)^2=\left(
\matrix{1 & 0 & 0 & 0 \cr 0 & \!\!\!\! -1 & 0  & 0 \cr 0 & 0 & \!\!\!\! -1 & 0
	\cr 0 & 0 & 0 & 1} \right)  \simeq Z_1  Z_2,
\]
where $Z_1=\left(R_{12}^{(6)}\right)^2$ and $Z_2=\left(R_{56}^{(6)}\right)^2$
are the Pauli $Z$-gates over the first and second qubits, respectively.
While the non-demolition measurement of the total topological  charge
might happen to be noisy, the implementation of the above quantum gate
as a 6-quasiholes state monodromy is completely  protected by topology.
\subsection{The two-qubit swap gate }
\label{sec:swap}
\begin{figure}[htb]
\centering
\includegraphics*[bb=0 260 570 560,width=10cm]{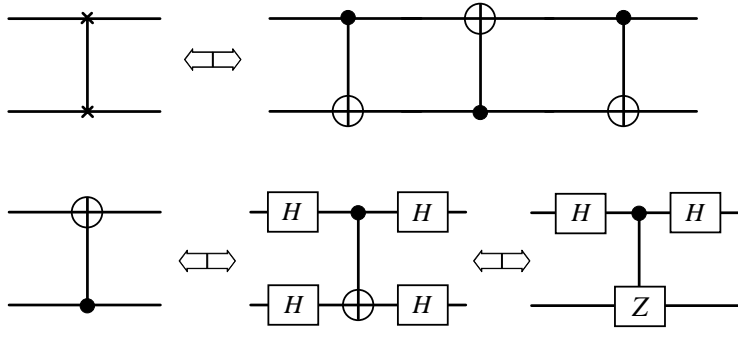}
\caption{The two-qubit SWAP gate realized by three CNOT's}
\label{fig:Swap}
\end{figure}

Once we now how to construct the CNOT gate it is straight forward to obtain the
two-qubit Swap gate in terms of three CNOTs \cite{nielsen-chuang}
as shown in the top line of Fig.~\ref{fig:Swap}. Here we
shall demonstrate that it is possible to implement the Swap gate with 15
elementary exchanges.
The bottom line of Fig.~\ref{fig:Swap}  shows how to express the swapped CNOT in terms of
the CZ and Hadamard gates. It turns out that in this circuit $H$ is essentially
equivalent to $R_{23}^{(4)}$
so that substituting $(H\otimes \I_2)\simeq  R_{23}^{(6)}$ and
$(\I_2\otimes H)\simeq  R_{45}^{(6)}$,
as well as using Eq.~(\ref{true-CZ}) for CZ, we finally   obtain
\beqa
\mathrm{Swap} & \simeq &  (\I_2 \otimes H) \ \mathrm{CZ} \ (\I_2 \otimes H) \ (H\otimes \I_2) \
\mathrm{CZ} \ (H\otimes \I_2) \  (\I_2\otimes H) \ \mathrm{CZ} \ (\I_2\otimes H) \nn
& \simeq &  R_{45} R_{34}^{-1} R_{12} R_{56} R_{45} R_{23}R_{34}^{-1} R_{12} R_{56}
  R_{23} R_{45} R_{34}^{-1} R_{12} R_{56} R_{45} \nn
 & = & i
 \left(
\matrix{1 & 0 & 0 & 0 \cr 0 & 0 & 1  & 0 \cr 0 & 1 & 0 & 0 \cr 0 & 0 & 0 & 1}
\right).
\eeqa
\section{Universal TQC scheme based on the Hadamard gate
	$H$, phase gate $S$, CNOT and Toffoli gate}
\label{sec:Toffoli}
\begin{figure}[htb]
\centering
\includegraphics*[bb=20 570 520 670,width=14cm]{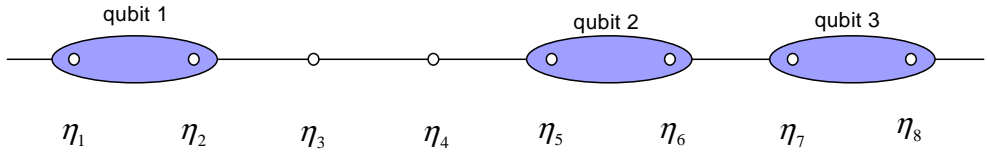}
 \caption{Three qubits constructed from 8 quasiholes}
\label{fig:3qubits}
\end{figure}

One of the standard universal set of gates, which could be used for
universal quantum computation, includes
the Hadamard gate $H$, the phase gate $S$, the two-qubit CNOT gate and
the Toffoli gate, which is a three-qubit  Controlled-Controlled-NOT
(CCNOT) gate \cite{nielsen-chuang}.
In order to execute three-qubit gates, such as the Toffoli
and Fredkin gates, we need to  consider a system with 8 quasiholes, whose
Hilbert subspace of states (for fixed positions of the quasiholes) has
dimension $2^{\frac{8}{2}-1}=8$. Here we shall assume that the third qubit is
defined by 2 more quasiholes at  positions $\eta_7$ and $\eta_8$
as shown on Fig.~\ref{fig:3qubits}. Then the three-qubit basis can be
written in terms of the Ising spin fields as
follows
\beqa \label{3qb-basis}
|000\ra &\equiv& \la \s_+ \s_+\s_+\s_+\s_+\s_+\s_+\s_+\ra, \quad
\! |001\ra \equiv \la \s_+\s_+\s_+ \s_-\s_+\s_+\s_+\s_-\ra, \nn
|010\ra &\equiv& \la \s_+\s_+\s_+ \s_-\s_+\s_-\s_+\s_+\ra, \quad
|011\ra \equiv \la \s_+\s_+\s_+ \s_+\s_+\s_-\s_+\s_-\ra, \nn
|100\ra &\equiv& \la \s_+\s_-\s_+ \s_-\s_+\s_+\s_+\s_+\ra, \quad
|101\ra \equiv \la \s_+\s_-\s_+ \s_+\s_+\s_+\s_+\s_-\ra, \nn
|110\ra &\equiv& \la \s_+\s_-\s_+ \s_+\s_+\s_-\s_+\s_+\ra, \quad
|111\ra \equiv \la \s_+\s_-\s_+ \s_-\s_+\s_-\s_+\s_-\ra  .
\eeqa
Now the fermion parity conservation requires that
\[
	e_3 e_4 =e_1 e_2 e_5 e_6 e_7 e_8,
\]
which reduces the number of independent states to from 16 to 8.
Although we have  chosen the quasiholes at $\eta_3$ and $\eta_4$ in such
a way to preserve the fermion parity, any other choice of the positions
of quasiholes representing the three
qubits would be equivalent to Eq.~(\ref{3qb-basis}) because the braid matrices
for the elementary exchanges would be just conjugated by an element of
the braid group, and the Artin relations (\ref{artin}) are invariant under
conjugation.
\subsection{Exchange matrices for 8 quasiholes}
Using the fusion rules (\ref{qh-OPE}) of the non-Abelian quasiholes we can express
the exchange matrices for 8 quasiholes recursively in terms of those for 6 quasiholes
as follows:
\beqa
R_{12}^{(8)}&=&\mathrm{diag}(1, 1, 1, 1, i, i, i, i)=R_{12}^{(6)}\otimes \I_2,
 \label{R8_12}\\
R_{23}^{(8)}&=&\frac{\e^{i\frac{\pi}{4}}}{\sqrt{2}}\left(
\matrix{1 & 0 & 0 & 0 & -i & 0 & 0 & 0 \cr 0 & 1 & 0 & 0 & 0 & -i  & 0 & 0 \cr
0 & 0 & 1 & 0 & 0 & 0  & -i & 0 \cr 0 & 0 & 0 & 1 & 0 & 0  & 0 & -i \cr
-i & 0 & 0 & 0 & 1 & 0 & 0 & 0 \cr 0 & -i & 0 & 0 & 0 & 1  & 0 & 0 \cr
0 & 0 & -i & 0 & 0 & 0   & 1 & 0 \cr 0 & 0 & 0 & -i & 0 & 0 & 0  & 1}
\right)=R_{23}^{(6)}\otimes \I_2,  \label{R8_23}\\
R_{34}^{(8)}&=&\mathrm{diag}(1,  i, i, 1, i, 1, 1, i),  \label{R8_34} \\
R_{45}^{(8)}&=&\frac{\e^{i\frac{\pi}{4}}}{\sqrt{2}}\left(
\matrix{1 & 0 & -i & 0 & 0 & 0 & 0 & 0 \cr 0 & 1 & 0 & -i & 0 & 0  & 0 & 0 \cr
-i & 0 & 1 & 0 & 0 & 0  & 0 & 0 \cr 0 & -i & 0 & 1 & 0 & 0  & 0 & 0 \cr
0 & 0 & 0 & 0 & 1 & 0 & -i & 0 \cr 0 & 0 & 0 & 0 & 0 & 1  & 0 & -i \cr
0 & 0 & 0 & 0 & -i & 0   &  1 & 0 \cr 0 & 0 & 0 & 0 & 0 & -i & 0  & 1}
\right)=R_{45}^{(6)}\otimes \I_2,  \label{R8_45}\\
R_{56}^{(8)}&=&\mathrm{diag}(1, 1, i, i, 1, 1, i, i)=R_{56}^{(6)}\otimes \I_2,
 \label{R8_56}\\
R_{67}^{(8)}&=&\frac{\e^{i\frac{\pi}{4}}}{\sqrt{2}}\left(
\matrix{1 & 0 & 0 & -i & 0 & 0 & 0 & 0 \cr 0 & 1 & -i & 0 & 0 & 0  & 0 & 0 \cr
0 & -i & 1 & 0 & 0 & 0  & 0 & 0 \cr -i & 0 & 0 & 1 & 0 & 0  & 0 & 0 \cr
0 & 0 & 0 & 0 & 1 & 0 & 0 & -i \cr 0 & 0 & 0 & 0 & 0 & 1  & -i & 0 \cr
0 & 0 & 0 & 0 & 0 & -i &  1 & 0 \cr 0 & 0 & 0 & 0 & -i & 0 & 0  & 1}
\right),  \label{R8_67} \\
R_{78}^{(8)}&=&\mathrm{diag}(1, i, 1, i, 1, i, 1, i)=\I_2 \otimes R_{56}^{(6)}
 \label{R8_78}.
\eeqa
It is not difficult to check that the exchange matrices (\ref{R8_12}),
(\ref{R8_23}), (\ref{R8_34}), (\ref{R8_45}), (\ref{R8_56}), (\ref{R8_67})
and (\ref{R8_78}) satisfy the Artin relations (\ref{artin}) for the braid group
$\B_8$.
Again, the order of the representation of the braid group $\B_8$ and its 
monodromy subgroup can be obtained by Dimino's algorithm \cite{dimino} to be
\[
|\mathrm{Image}(\B_8)|= 5160960, \quad |\mathrm{Image}(\mathcal{M}_8)|=128.
\]
As an illustration of the derivation of the 8-quasiholes exchange matrices,
let us compute the last row of $R_{67}^{(8)}$, i.e., we consider the 
transformation
of the state $|111\ra$ when we exchange $\eta_6$ with $\eta_7$.
Because the state of the second and the third qubits is independent of
the quasiholes at $\eta_1$ and $\eta_2$, we could fuse $\eta_1\to \eta_2$ obtaining in
this way a 6-quasiholes state whose braiding properties are already known.
Indeed, using the OPE (\ref{qh-OPE}) we find
\beqa
 |111\ra &\mathop{\simeq}_{\eta_1\to \eta_2}&  \sqrt{\frac{\eta_{12}}{2}}
 \la \psi(\eta_2)\s_+(\eta_3)\s_-(\eta_4) \s_+(\eta_5) \s_-(\eta_6) \s_+(\eta_7)\s_-(\eta_8)\ra
 \nn
& \mathop{\simeq}_{\eta_2\to \eta_3} & \sqrt{\frac{\eta_{12}}{2}} \sqrt{\frac{1}{2\eta_{23}}}
 \la \s_-(\eta_3)\s_-(\eta_4) \s_+(\eta_5) \s_-(\eta_6) \s_+(\eta_7)\s_-(\eta_8)\ra \simeq
 |01\ra \nonumber
 \eeqa
where we used the OPE \cite{fst}
$\psi(\eta_2)\s_{e_3}(\eta_3) \simeq (2\eta_{23})^{-1/2}\s_{-e_3}(\eta_3)$,
for $\eta_2 \to \eta_3$, and the identity
$\la \s_-\s_-\s_+ \s_- \s_+\s_-\ra \simeq \la \s_+\s_+\s_+ \s_- \s_+\s_-\ra \equiv |01\ra$.
It is now obvious that the exchange of $\eta_{6}$ with $\eta_7$ in the three-qubit state
$|111\ra$ is equivalent to the exchange of the fifth and sixth quasiholes in the state
$|01\ra$ so that, taking  $R_{45}^{(6)}$ from Eq.~(\ref{R6_45}), we obtain
 \beqa
 |111\ra 	&\mathop{\rightarrow}\limits^{R_{45}^{(6)}}&
 \frac{\e^{i\frac{\pi}{4}}}{\sqrt{2}} \sqrt{\frac{\eta_{12}}{2}}
 \sqrt{\frac{1}{2\eta_{23}}}
\left(-i |00\ra +|01\ra \right)    \nn
 &=&\frac{\e^{i\frac{\pi}{4}}}{\sqrt{2}}
 \sqrt{\frac{\eta_{12}}{2}} \sqrt{\frac{1}{2\eta_{23}}}
\left(-i
 \la \s_+(\eta_3)\s_+(\eta_4)\s_+(\eta_5)\s_+(\eta_6)\s_+(\eta_7)\s_+(\eta_8)\ra \right. \nn
 &+&\left. \la \s_+(\eta_3)\s_+(\eta_4)\s_+(\eta_5)\s_-(\eta_6)\s_+(\eta_7)\s_-(\eta_8)\ra
 \right)\nn
 &\mathop{\simeq}_{\eta_2\to \eta_3} &
   \frac{\e^{i\frac{\pi}{4}}}{\sqrt{2}}  \sqrt{\frac{\eta_{12}}{2}}
\left(-i
 \la \psi(\eta_2)\s_-(\eta_3)\s_+(\eta_4)\s_+(\eta_5)\s_+(\eta_6)\s_+(\eta_7)\s_+(\eta_8)\ra \right. \nn
 &+&\left. \la \psi(\eta_2)\s_-(\eta_3)\s_+(\eta_4)\s_+(\eta_5)\s_-(\eta_6)\s_+(\eta_7)\s_-(\eta_8)\ra
 \right)\nn
 &\mathop{\simeq}_{\eta_1\to \eta_2} &
   \frac{\e^{i\frac{\pi}{4}}}{\sqrt{2}} \left(-i
 \la \s_+(\eta_1)\s_-(\eta_2)\s_+(\eta_3)\s_-(\eta_4)\s_+(\eta_5)\s_+(\eta_6)\s_+(\eta_7)\s_+(\eta_8)\ra \right. \nn
 &+&\left. \la \s_+(\eta_1)\s_-(\eta_2)\s_+(\eta_3)\s_-(\eta_4)\s_+(\eta_5)\s_-(\eta_6)\s_+(\eta_7)\s_-(\eta_8)\ra
 \right)\nn
   &=&\frac{\e^{i\frac{\pi}{4}}}{\sqrt{2}} \left(-i |100\ra +|111\ra \right) ,
\eeqa
which exactly reproduces the last row of $R_{67}^{(8)}$. In the above derivation
we restored the 8-quasiholes states using the same OPEs for $\eta_2\to \eta_3$ and
 $\eta_1\to \eta_2$, however in reverse,
as well as used the identity 
$\s_-(\eta_3)\s_+(\eta_4) \simeq \s_+(\eta_3)\s_-(\eta_4)$.

\begin{rem}
Due to the specifics of  the braid group representation,
it may not be always possible to represent exactly the single- and
two- qubit gates in the three-qubit basis (\ref{3qb-basis}). Indeed, the
6-quasiholes exchange matrix $R_{34}^{(6)}$, defined in Eq.~(\ref{R6_34}),
is not a factorized tensor product of the exchange matrices for 4 quasiholes,
rather it contains the built-in CZ matrix (\ref{true-CZ}). Therefore some 
tensor products, which are trivial otherwise, might not be constructed 
directly in terms of the elementary exchange matrices for 8 quasiholes.
One consequence of this peculiarity is that some one-qubit and two-qubit 
gates would be easier realizable in the three-qubit basis (\ref{3qb-basis}) 
in terms of elementary 8-quasiholes exchange matrices, however, up to a pair 
of extra minus signs, see Sect.~\ref{sec:embedding}. While the exact 
construction of the one-qubit and two-qubit gates would require more work, 
their simplified versions might be sufficient in most cases.
\end{rem}
\subsection{Embedding of one-qubit and two-qubit gates into a three-qubit
	system}
	\label{sec:embedding}
The three one-qubit phase gates are directly expressed as single elementary
8-quasiholes exchange matrices, i.e.,
\[
 S_1\equiv S\otimes \I_4 = R_{12}^{(8)}, \quad
 S_2\equiv  \I_2\otimes S\otimes \I_2 =  R_{56}^{(8)}, \quad
   S_3\equiv \I_4\otimes S = R_{78}^{(8)}.
\]
The first one-qubit Hadamard gate can be constructed exactly in terms of the exchange
matrices for 8 quasiholes  by (skipping the superscript ``$(8)$'' of $R$)
\beqa
H_1=H\otimes \I_4 \simeq R_{12}^{-1}R_{23}^{-1}R_{12}^{-1}
 &=& \frac{\e^{-i\frac{\pi}{4}}}{\sqrt{2}}\left(
\matrix{1 & 0 & 0 & 0 & 1 & 0 & 0 & 0 \cr 0 & 1 & 0 & 0 & 0 & 1  & 0 & 0 \cr
0 & 0 & 1 & 0 & 0 & 0  & 1 & 0 \cr 0 & 0 & 0 & 1 & 0 & 0  & 0 & 1 \cr
1 & 0 & 0 & 0 & \! \! \! \! -1 & 0 & 0 & 0 \cr 0 & 1 & 0 & 0 & 0 & \! \! \! \! -1  & 0 & 0 \cr
0 & 0 & 1 & 0 & 0 & 0   &  \! \! \! \! -1 & 0 \cr 0 & 0 & 0 & 1 & 0 & 0 & 0  & \! \! \! \! -1}
\right),
\eeqa
while the second Hadamard gate could be constructed as follows
\beq
H_2 \simeq \I_2 \otimes H \otimes \I_2 \simeq
R_{56}^{-1}  R_{45}^{-1}R_{56}^{-1} = \frac{\e^{-i\frac{\pi}{4}}}{\sqrt{2}}\left(\matrix{
	1& 0& 1& 0& 0& 0& 0& 0 \cr 0& 1& 0& 1& 0& 0& 0& 0 \cr
	1& 0& \! \! \! \! -1& 0& 0& 0& 0& 0 \cr
	0& 1& 0& \! \! \! \! -1& 0& 0& 0& 0 \cr
	0& 0& 0& 0& 1& 0& 1& 0 \cr 0& 0& 0& 0& 0& 1& 0& 1 \cr
	0& 0& 0& 0& 1& 0& \! \! \! \! -1& 0 \cr
	0& 0& 0& 0& 0& 1& 0& \! \! \! \! -1}\right) . \quad
\eeq
The third Hadamard gate could also be reproduced up to some swapping by
\beqa \label{H_3}
H_3 &=&\I_4\otimes H \simeq
R_{78}^{-1}R_{45}^{-1}R_{56}^{-1}R_{67}^{-1}R_{56}^{-1}R_{45}^{-1}R_{78}^{-1}
 \nn
&=& \frac{-\e^{i\frac{\pi}{4}}}{\sqrt{2}} \left( \matrix{
1 & 1 & 0 & 0 & 0 & 0 & 0 & 0 \cr
1 & \! \! \! \! - 1 & 0 & 0 & 0 & 0  & 0 & 0 \cr
0 & 0 & \! \! \! \! -1 & 1 & 0 & 0  & 0 & 0 \cr
0 & 0 & 1 & 1 & 0 & 0  & 0 & 0 \cr
0 & 0 & 0 & 0 &  1 & 1 & 0 & 0 \cr
0 & 0 & 0 & 0 &  1 & \! \! \! \! -1  & 0 & 0 \cr
0 & 0 & 0 & 0 & 0 & 0   &  \! \! \! \! -1 & 1 \cr
0 & 0 & 0 & 0 & 0 & 0 & 1 & 1}
\right) .
\eeqa
It might be useful to give also the realization of the single-qubit NOT gates:
$X_1 \equiv X\otimes \I_4=R_{23}^2$, 
$X_2\equiv\I_2\otimes X \otimes \I_2=R_{45}^2$ and
$X_{3}\equiv\I_4\otimes X = R_{45}^2 R_{67}^2$.

We should stress again that the topological entanglement between the qubits
mentioned in Sect.~\ref{sec:single-qubit} leads to serious difficulties for 
efficient embedding of
the one-qubit and two-qubit gates in three-qubit systems.
For example, the NOT gate $X_3$ acting on the third qubit has a different 
structure than just a tensor product of the exchange matrix producing the 
single-qubit $X$ gate. Similarly, the Hadamard gates acting on the first 
and second qubits have slightly different structures from their single-qubit 
counterpart $H$, while that  acting on the third qubit
cannot  even be obtained exactly with the same number of elementary exchanges 
as $H$.
This seems to be a common problem arising in all TQC schemes using 
non-Abelian anyons in FQH systems, whose general solution is still missing.
Moreover, it appears that the two-qubit Controlled-NOT gates in a three-qubit 
systems cannot be 
directly constructed in the three-qubit basis (\ref{3qb-basis}) because of 
the topological entanglement between the two qubits and the third one.
This requires more work and will be reported elsewhere.
Just for reference, we give a simple  implementation of a three-qubit 
operation which is very close to 
$\mathrm{CNOT}_2=\I_2\otimes \mathrm{CNOT}$ 
\beqa \label{CNOT_2}
\widetilde{\mathrm{CNOT}}_2 &\simeq&{\I_2\otimes \mathrm{CNOT}}\simeq
R_{12}^{-1}R_{56}\widetilde{\mathrm{SWAP}}_2 R_{36} R_{45}
\widetilde{\mathrm{SWAP}}_2 \nn
&=& \left(
\matrix{1 & 0 & 0 & 0 & 0 & 0 & 0 & 0 \cr 0 & 1 & 0 & 0 & 0 & 0  & 0 & 0 \cr
0 & 0 & 0 & 1 & 0 & 0  & 0 & 0 \cr 0 & 0 & 1 & 0 & 0 & 0  & 0 & 0 \cr
0 & 0 & 0 & 0 & 0 & 1 & 0 & 0 \cr 0 & 0 & 0 & 0 & 1 & 0  & 0 & 0 \cr
0 & 0 & 0 & 0 & 0 & 0  & 1 & 0 \cr 0 & 0 & 0 & 0 & 0 & 0  & 0 & 1}
\right),
\eeqa
where $R_{36}=R_{56}^{-1}R_{45}^{-1}R_{34}^{-1}R_{45}R_{56}$ and the 
gate $\widetilde{\mathrm{SWAP}}_2$ is defined below.

The two-qubit SWAP gates can be simply realized by braiding in the 
three-qubit basis (\ref{3qb-basis}) (up to overall phases and pairs of 
extra minus signs) by
\beqa
\widetilde{\mathrm{SWAP}}_1 &\simeq& \mathrm{SWAP}\otimes \I_2\simeq
R_{45}R_{56}^{-1}R_{34}^{-1} R_{45} R_{23}R_{34}^{-1}R_{12}^{-1}R_{23}
R_{45}R_{56}^{-1}R_{34}^{-1} R_{45}  \nn
&=& \left( \matrix{-1& 0& 0& 0& 0& 0& 0& 0 \cr 0& -1& 0& 0& 0& 0& 0& 0 \cr
	0& 0& 0& 0& 1& 0& 0& 0 \cr 0& 0& 0& 0& 0& -1& 0& 0 \cr 0& 0& 1& 0& 0& 0& 0& 0 \cr
	0& 0& 0& -1& 0& 0& 0& 0 \cr 0& 0& 0& 0& 0& 0& 1& 0 \cr 0& 0& 0& 0& 0& 0& 0& 1} \right)
\eeqa
\[
\widetilde{\mathrm{SWAP}}_2\simeq
\I_2 \otimes \mathrm{SWAP}\simeq R_{67} R_{78}^{-1}R_{56}^{-1}R_{67}=
\left(\matrix{-1& 0& 0& 0& 0& 0& 0& 0 \cr 0& 0& 1& 0& 0& 0& 0& 0 \cr
0& 1& 0& 0& 0& 0& 0& 0 \cr 0& 0& 0& 1& 0& 0& 0& 0 \cr 0& 0& 0& 0& -1& 0& 0& 0 \cr
0& 0& 0& 0& 0& 0& 1& 0 \cr 0& 0& 0& 0& 0& 1& 0& 0 \cr 0& 0& 0& 0& 0& 0& 0& 1} \right)
\]
The Swap gates turn out to be very important because they can be used to
construct gates acting on one of the qubits in terms of similar gates acting on another qubit.
For example, the CNOT gate acting on the first and the third qubits can be expressed
in terms of the CNOT acting on the first and the second qubit plus two gates
$\mathrm{SWAP}_2$.
\begin{figure}[htb]
\centering
\includegraphics*[bb=0 400 530 620,width=8cm]{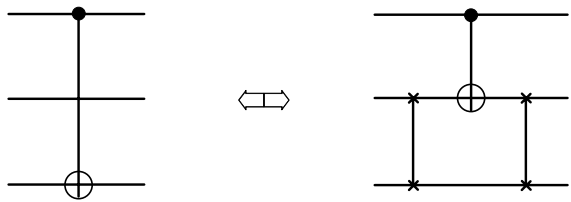}
 \caption{The CNOT acting on qubits 1 and 3 expressed in terms of
 	$\mathrm{CNOT}\otimes \I_2$, which is the CNOT gate acting on
 	qubits 1 and 2, and two   gates $\I_2\otimes \mathrm{SWAP}$.}
\label{fig:CNOT13}
\end{figure}
The extra minus signs appearing in some of the three-qubit 
operations are not an innocent thing, because they may have different 
properties from the standard gates. However, it appears that in many cases 
these simplified gates, which are much simpler to construct,  can be used 
instead of the standard once.

The three-qubit Toffoli gate \cite{nielsen-chuang} can be constructed in terms of 
the Controlled-$S$ gate and CNOT like in Ref.~\cite{TQC-PRL} using the relation 
between the Toffoli gate and the 
Controlled-Controlled-Z gate or by a braid-group 
based Controlled-Controlled-$Z$ gate  precursor Ref.~\cite{TQC-PRL}, which must be
supplemented by the Bravyi--Kitaev construction \cite{TQC-PRL}.
\section{Discussion}
In this paper we explicitly implemented all single-qubit gates in the 
Pfaffian TQC scheme, except for the $\pi/8$ one, in terms of 4-quasihole 
braidings, as well as the two-qubit Controlled-Z and CNOT gates in terms 
of 6-quasihole braidings.
These gates, which are known to form a Clifford group, are realized in a
completely topologically protected way because of the topological nature of 
the braid operations in the FQH liquids.
This work is an extension of the topological quantum computation scheme of 
Ref.~\cite{sarma-freedman-nayak} using pairs of Pfaffian quasiholes localized 
on  antidots to construct elementary qubits and execute logical NOT on 
them.
While the original TQC scheme of Ref.~\cite{sarma-freedman-nayak} used
only monodromy transformations to realize quantum gates, we, for the first 
time,  exploited explicitly quasihole braiding in the Pfaffian FQH state 
to construct   
 the single-qubit Hadamard gate $H$, the phase gate $S$ and the CNOT gate.
Although the Gottesmann--Knill theorem
says that any circuit based only on the Clifford group gates could be
efficiently simulated on a (probabilistic) classical computer these gates
do  play a crucial role in quantum computation, especially in the 
error-correcting algorithms \cite{nielsen-chuang}.

Due to the topological entanglement between the separate qubits realized by 
non-Abelian anyons in FQH systems some difficulties arise when trying to 
embed the one-qubit and two-qubit gates into systems with more qubits.
This makes the embedding of Clifford gates non-trivial and requires more 
work.

For implementing three-qubit gates such as the Toffoli and Fredkin gates
\cite{nielsen-chuang}, in the Pfaffian TQC scheme,  we considered
Pfaffian wave functions with  8-quasiholes, in which case the topological
degeneracy of the space of correlation functions is $8$ 
\cite{mr,nayak-wilczek}). We derived explicitly the braid 
matrices for the elementary 8-quasiholes exchanges, which
serve as building blocks for constructing 
all three-qubit gates. More work in this direction, including eventually 
the topologically protected construction of the  Toffoli gate would be  
reported elsewhere.

To conclude, let us make  some remarks about the possible observation of
the non-Abelian statistics. We believe that in order to observe the 
Pfaffian phase at $\nu=5/2$
one should perform the experiment at temperature below 15 mK.
The point is that there might exists another incompressible Abelian phase,
which was called the Extended Pfaffian (EPf) state  \cite{5-2},
 that could also be realized at $\nu=5/2$. Perhaps the most observable
difference between the two phases is
in the electric charge of their quasiparticles: $1/4$ for the Pfaffian
and $1/2$ for the EPf.
The EPf phase was obtained mathematically by a local chiral algebra extension
of the Pfaffian state and satisfies all conditions necessary for an
incompressible quantum Hall state \cite{5-2}.
The motivation for introducing this
new state is that there is a persistent unexplained kink around $T_c=15$ mK
observed in the thermal activation experiment \cite{pan}
  showing two different slops that presumably correspond to two different
gaps below and above the critical temperature.
Analyzing the edge states CFT it has been proposed in Ref.~ \cite{5-2}
a possible explanation of the
kink in terms of a finite temperature two-step phase transition between
the Pfaffian and the EPf state involving an intermediate compressible
state of composite fermions. Here is a brief description of the
process (see Sect. 9 in Ref.~\cite{5-2} for more
details):
at low temperature the system is definitely in the Pfaffian phase,
as the numerical calculations suggest. As temperature increases to
about $1/2$ of the Pfaffian-phase gap, which was estimated to be
about 33 mK (for the sample of Ref.~\cite{pan}),
the system becomes more and more
compressible (look at the behavior of the free energy on the edge,
Fig. 5 in Ref.~\cite{5-2}) leading to a II-nd
order phase transition to the compressible state of composite fermions
(which has the same topological structure like the Pfaffian state but having
at the same time the $\Z_2$ symmetry of the EPf state that is broken
spontaneously in the Pfaffian phase). Immediately after that, as temperature
continues to increase, there is
a I-st order phase transition to the EPF state
which is expected to have a higher gap than the Pfaffian state.

\begin{ack}
I thank Ivan Todorov, Ady Stern,  Valentina Petkova, Chetan Nayak,
Lyudmil Hadjiivanov and Michael Geller for many helpful discussions.
This work has been partially supported by the FP5-EUCLID Network Program
 of the European Commission under Contract No. HPRN-CT-2002-00325
and by the Bulgarian National
Council for Scientific Research under Contract No. F-1406.
\end{ack}

\appendix
\section{Binomial series expansion of the 4-pt function and analytic
	continuation}
	\label{app:Laurent}
\begin{figure}[htb]
\centering
\includegraphics*[bb=0 340 580 770,width=\textwidth]{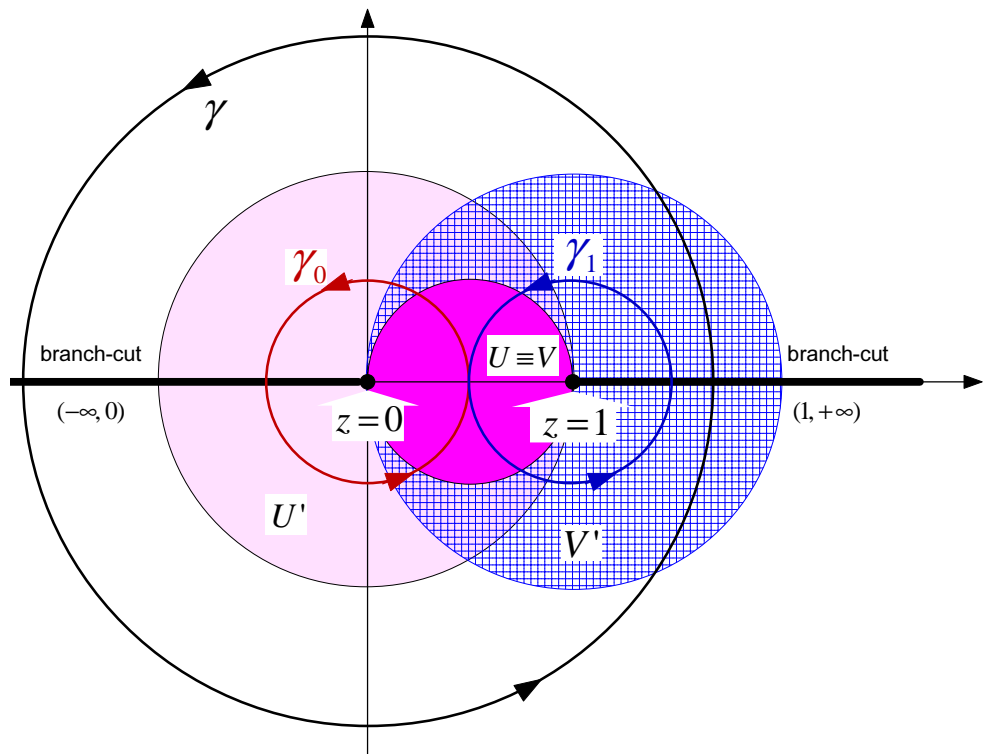}
 \caption{(Color online). Domains and contours for the different
 	values of $|z|$ used for
 	the analytic continuation and binomial series expansion}
\label{fig:domain}
\end{figure}

In this appendix we shall give some details  about the analytic
continuation of the function (\ref{f}), which might be necessary
for the understanding of the results of Sect.~\ref{sec:read-out}.
Using the standard complex analysis notion  \cite{shabat}
of a branching point as a multi-valued  isolated singular point,
we consider a punctured neighborhood of the branching point, denoted as $U'$
in which we would like to continue the element $(U,f)$ of the function
 (\ref{f}) from the simply-connected sub-domain $U\subset U'$ along
any path. For example, for the branching point at $z=0$ we can choose
\[
U'=\{ z\  | \ 0< |z|< 1\}, \quad
U=\left \{ z \ | \ \left|z-1/2 \right|< 1/2 \right \},
\]
as shown on Fig.~\ref{fig:domain}.
Then the analytic continuation  along the contour
\beq \label{gamma_0}
\gamma_0= \left \{ z \ \Big| \ z=\frac{1}{2}\e^{it},
\quad 0\leq t \leq 2\pi \right \}
\eeq
changes the sign in front of the inner root but not the one of the outer
root because
$\gamma_0$ does not encircle $z=1$. This can be verified directly
by using the (fractional powers) Laurent expansion, which in this case
can be obtained by the binomial series expansion
\[
(1+x)^\alpha = \sum_{n=0}^\infty
\frac{\Gamma(\alpha+1)}{\Gamma(n+1)\Gamma(\alpha+1-n)} x^n , \quad |x|<1
\]
applied for $\alpha=1/2$. Using the defining  $\Gamma$ function property
$\Gamma(z)=(z-1)\Gamma(z-1)$, we get
\beq \label{z-0}
\sqrt{1\pm \sqrt{z}}=\left(1\pm\frac{1}{2}\sqrt{z} - \frac{1}{8} z \pm
\frac{1}{16}z \sqrt{z} - \cdots \right), \quad z \in U' .
\eeq
That is why  $\sqrt{1 \pm \sqrt{z}} \to \sqrt{1 \mp \sqrt{z}} $,   for $|z|<1$,
when continuing $z\to \e^{2\pi i}z$ along any contour in $U'$, which is
homotopic to
(\ref{gamma_0}).

For the branching point at $z=1$, on the other hand, we consider  the
domains $V'$ and $V$ shown again on Fig.~\ref{fig:domain}  defined by
\[
V'=\{ z\  | \ 0< |z-1|< 1\}, \quad
	 V=\left \{ z \ | \ \left|z-1/2 \right|< 1/2 \right \}.
 \]
Then the analytic continuation of the element $(V,f)$ of the function
 (\ref{f}) from $V$ to $V'$ along any contour in $V'$ homotopic to
 \[
	 \gamma_1= \left \{ z \ \Big| \ z=1+\frac{1}{2}\e^{it},
	 	\quad 0\leq t \leq 2\pi \right \}
 \]
does not change the inner root sign because the point $z=0$ is outside
the contour. It only changes the sign of the outer root whenever the
inner root sign is ``$-$''. Indeed,
representing
$\sqrt{z}=\sqrt{1+(z-1)}\simeq \left(1+\frac{z-1}{2}- \frac{(z-1)^2}{8}+
\frac{(z-1)^3}{16}\right)$, for $|z-1|\ll 1$, and using again
the binomial expansion in $V'$ we get
\beqa
\sqrt{1+\sqrt{z}} &=& \sqrt{2}\left(1+ \frac{z-1}{8} - \frac{5}{128}(z-1)^2 +
\cdots\right) \label{z+1}\\
\sqrt{1-\sqrt{z}} &=& \frac{\sqrt{z-1}}{i\sqrt{2}}\left(1- \frac{z-1}{8} +
\frac{7}{128}(z-1)^2 +\cdots\right) , \quad z \in V' . \label{z-1}
\eeqa
The appearance of $\sqrt{z-1}$ in the right-hand side  of
Eq.~(\ref{z-1}) but not in
Eq.~(\ref{z+1})
implies that when $z$ encircles the point $1$ the function
$\sqrt{1-\sqrt{z}}$ acquires one minus sign, while the function
$\sqrt{1+\sqrt{z}}$ is single-valued.

Recall that for a contour passing through the points $z=0$ or $z=1$ it is not
possible to make analytic continuation.

Finally for $|z|>1$ we use
$W'=\{ z\  | \ 1< |z|< \infty \}, \quad
	 W =\left \{ z \ | \ \left|z-2 \right|< 1 \right \}$,
and the continuation  along the contour
 \[
	 \gamma_2= \left \{ z \ \Big| \ z=2\e^{it},
	 	\quad 0\leq t \leq 2\pi \right \}
 \]
 changes both the sing of the inner root and that of
 the outer root when the inner sign is ``$-$''.
Because in this case $|1/\sqrt{z}|<1$ the binomial expansion
with respect to $1/\sqrt{z}$ gives
\[
\sqrt{1\pm \sqrt{z}} = \sqrt[4]{z}
\left(1\pm \frac{1}{2\sqrt{z}} - \frac{1}{8z} \pm
\frac{1}{16z\sqrt{z}}  	\cdots\right) , \quad 1<|z|<\infty
\]
and that explicitly shows that $z=\infty$  is a branching point of order $4$.

That is how we conclude that  the general  contour $\gamma$, which i shown
on Fig.~\ref{fig:domain}, corresponding to the
read-out transformation
(\ref{read-out}) is homotopic to $\gamma_0$ or $\gamma_0  \cup \gamma_1 $
depending on the value of $|x|$, i.e.,
\[
\gamma \simeq \left\{
	\begin{array}{ll}\gamma_0 & \quad \mathrm{for} \quad |x|<1 \\
	 \gamma_0 \cup \gamma_1 & \quad \mathrm{for} \quad |x|>1
	 \end{array}\right.  ,
\]
 which explains once again the analytic continuation
  results obtained in Sect.~\ref{sec:read-out}.
\bibliography{FQHE,Z_k,my,TQC}

\end{document}